\documentclass[12pt,preprint]{aastex}
\usepackage{amsmath,amssymb}
\usepackage{hyperref}
\let\captionbox\relax
\usepackage{graphicx,caption,subcaption}
\usepackage{natbib}
\usepackage{multirow}
\begin{document}
\title{Magnetohydrodynamic simulation of magnetic null-point reconnections in NOAA AR12192 initiated with an extrapolated non-force-free-field}

\author{A. Prasad, R. Bhattacharyya} 
\affil{Udaipur Solar Observatory, Physical Research Laboratory, Dewali, Bari Road, Udaipur-313001, India}

\author{Qiang Hu} 
 \affil{Department of Space Science and Center for Space Plasma and Aeronomic Research, The University of Alabama in Huntsville, Huntsville, AL 35899, USA}
 
\author{Sanjay Kumar}
\affil{Post Graduate Department of Physics,
Patna University, Patna-800005, India}

\author{Sushree S. Nayak}
\affil{Udaipur Solar Observatory, Physical Research Laboratory, Dewali, Bari Road, Udaipur-313001, India}

\begin{abstract}
Magnetohydrodynamics of the solar corona is simulated numerically. The simulation is initialized with an extrapolated non-force-free magnetic field using the vector magnetogram of the active region (AR) NOAA 12192 obtained on the solar photosphere. Particularly, we focus on the magnetic reconnections occurring close to a magnetic null-point that resulted in appearance of circular chromospheric flare ribbons on October 24, 2014 around 21:21 UT, after peak of an X3.1 flare.  The  extrapolated field lines show the presence of the three-dimensional (3D) null near one of the polarity inversion lines---where the flare was observed. In the subsequent numerical simulation, we find magnetic reconnections  occurring near the null point, where the magnetic field lines from the fan-plane of the 3D null form a X-type configuration with underlying arcade field lines. The footpoints of the dome-shaped field lines, inherent to the 3D null, show high gradients of the squashing factor. We find slipping reconnections at these quasi-separatrix layers, which are co-located with the post-flare circular brightening observed at the chromospheric heights. This demonstrates the viability of the initial non-force-free field along with the dynamics it initiates. Moreover, the initial field and its simulated evolution is found to be devoid of any flux rope, which is in congruence with the confined nature of the flare.
\end{abstract}

\keywords{magnetohydrodynamics (MHD) -- Sun: activity -- Sun: corona -- Sun: flares -- Sun: magnetic fields -- Sun: photosphere}

\section{Introduction}
\label{sec:1}

The solar corona can be treated as a magnetized plasma having large electrical conductivity with evolution being determined by the magnetohydrodynamic (MHD) equations {\citep{2014masu.book.....P}}. The magnetic Reynolds number $R_M (vL/\eta$, in usual notations) for the corona is of the order of $10^{10}$ {\citep{2004psci.book.....A}}, which 
makes the Alfv\'{e}n's  theorem of flux freezing valid and ensures plasma-parcels to remain tied with magnetic field lines (MFLs) during evolution {\citep{1942Natur.150..405A}}. The eruptive events (flares, coronal mass ejections) occurring at the  corona are thought to be signatures of magnetic reconnection (MR): a process involving the topological rearrangement of MFLs with conversion of  magnetic energy into heat and kinetic energy of mass motion {\citep{2011LRSP....8....6S}}. Notably, the requirement to onset MRs is small $R_M$ which corresponds to small $L$, the length over which the magnetic field varies.  The smallness of  $L$ can either be pre-existing in a magnetic topology---manifested as magnetic nulls and quasi-separatrix layers (QSLs)---or can develop autonomously during the evolution of the magnetofluid. Such autonomous developments (owing to discontinuities in magnetic field) are expected from the Parker's magnetostatic theorem \citep{1972ApJ...174..499P,1988ApJ...330..474P,1994ISAA....1.....P} which states that for a perfect electrically conducting plasma, the conditions of flux-freezing and the equilibrium cannot be satisfied simultaneously by a magnetic field which is continuous everywhere. The reduction of $L$ and the consequent spontaneous magnetic reconnections during a quasi-static evolution of the plasma under a near-precise maintenance of the flux-freezing has been identified in contemporary MHD simulations {\citep{2015PhPl...22a2902K,2016PhPl...23d4501K,2016ApJ...830...80K}}. However, these studies were performed using idealized scenarios of initial bipolar magnetic fields that lacked  the complexities often observed in  solar active regions. 

Presently, the coronal field needs to be extrapolated from the photospheric magnetic field because of a lack of direct measurements. 
For extrapolation, the usage of the nonlinear-force-free-fields(NLFFFs), a subset of force-free-fields \citep{2008JGRA..113.3S02W,2012LRSP....9....5W} is customary. The NLFFF can be solved analytically (in spherical polar coordinates) under the assumption of axisymmetry  \citep{1990ApJ...352..343L,2014ApJ...786...81P} but the analytical solution fails to effectively capture the complexity of an active region magnetogram, which is often non-axisymmetric. Such complexities are well replicated in NLFFF extrapolations \citep{2017ApJ...842..119D}. Recent MHD simulations based on 
NLFFF extrapolations were successful in simulating  the coronal dynamics leading to eruptions \citep{2013ApJ...771L..30J, 2013ApJ...779..129K, 2014Natur.514..465A, 2014ApJ...788..182I, 2015ApJ...803...73I, 2015ApJ...810...96S, 2016ApJ...817...43S,2016PEPS....3...19I}. Importantly, only the region sandwiched between the photosphere and the upper corona
is relatively force-free  whereas at the photosphere---where magnetograms are obtained---the Lorentz force is non-zero \citep{2001SoPh..203...71G}. 
Generally, to mitigate this problem within the framework of NLFFF, a technique called `preprocessing' is often performed on the photospheric data which minimizes the Lorentz 
force in the vector magnetograms and provides a boundary condition suitable for NLFFF extrapolations \citep{2006SoPh..233..215W,2014SoPh..289...63J}.

An alternative is the extrapolation using non-force-free-fields (NFFFs) described by the double-curl Beltrami equation for the magnetic field ${\bf{B}}$ \citep{2008SoPh..247...87H}. The equation has been analytically solved for an idealized corona  \citep{2007SoPh..240...63B, 2011PhPl...18h4506K}
to obtain MFLs resembling coronal loops. 
Recently, a semi-analytical construction based on maximizing correlations of non-axisymmetric NFFFs with photospheric vector magnetograms of NOAA AR11283 
successfully mimicked an event of filament bifurcation by tracking MHD evolution of a pre-existing flux -rope \citep{2016PhPl...23k4504P,2017ApJ...840...37P}
However, missing from the simulation were the small scale magnetic features and their influence on the MFL dynamics---which cannot be captured
by analytical/semi-analytical models. 
To include these magnetic features and determine their role in overall magnetofluid evolution, 
here we numerically simulate evolution of AR 12192 initiated with the NFFF extrapolation model 
developed by \citet{2008SoPh..247...87H, 2008ApJ...679..848H,2009SoPh..257..271G}. 
The focus is to assess the viability of the NFFF extrapolation 
in generating MFLs, the evolution of which can reliably imitate the observed dynamics.
For the purpose, we follow the evolution of AR 12192, starting at 20:46 UT on 24th October 2014 and study the X3.1 confined flare occurring at 21:10 UT.  Importantly, a non-zero Lorentz force, instead of prescribed flows  \citep{2003ApJ...585.1073A,2010ApJ...708..314A}, is envisaged here to initiate dynamics.

In the rest of the paper, Section 2 discusses the flare-event and the observations required for the NFFF extrapolation. In Section 3, we present the details of the initial extrapolated field. The MHD model is discussed in Section 4. The results of the simulation are presented in Section 5 and the Section 6 summarizes important results.

\section{Discussions on the X3.1 flare event}
\label{sec:2}

The AR12192 was the largest of all active regions appearing in the solar cycle 24 which produced a series of X-class flares \citep{2015ApJ...808L..24C}. The X3.1 flare on October 24, 2014 around 21:15 UT was the strongest in a series which did not lead to any coronal mass ejection (CME) \citep{2015ApJ...804L..28S,2018SoPh..293...16S}. Since there is a very strong correlation between flare intensity and occurrence of CMEs \citep{2005JGRA..11012S05Y}, this event has been extensively studied. An absence of flux rope was suggested in \citet{2016ApJ...828...62J} for explaining the confined nature whereas the onset of the flare was attributed to tether-cutting (TC) MRs  \citep{2001ApJ...552..833M} between sheared arcades. Further studies of successive strong X-class flares triggered by TC reconnections, in the same AR, were also reported in \citet{2015ApJ...808L..24C}. Contrarily,  using NLFFF extrapolation,  \citet{2016ApJ...818..168I} found a multiple-flux-tube system located near a Polarity Inversion Line (PIL) to be favorable for  the TC reconnections. They attributed the stability of the flux-tube-system to the  overlying strong tethering MFLs.  Similar results were also documented in \citet{2015ApJ...808L..24C}, where the mean decay index of the horizontal background field was found to be less than the typical threshold required for the torus instability \citep{2006PhRvL..96y5002K} to set in. An alternative explanation was provided  by \citet{2017ApJ...845...54Z} who attributed the confined nature 
to the complexity of the involved magnetic field structures.

\begin{figure}[hp]
  \centering
  \begin{subfigure}[]{0.7\textwidth}
    \centering
    \includegraphics[width=1\linewidth]{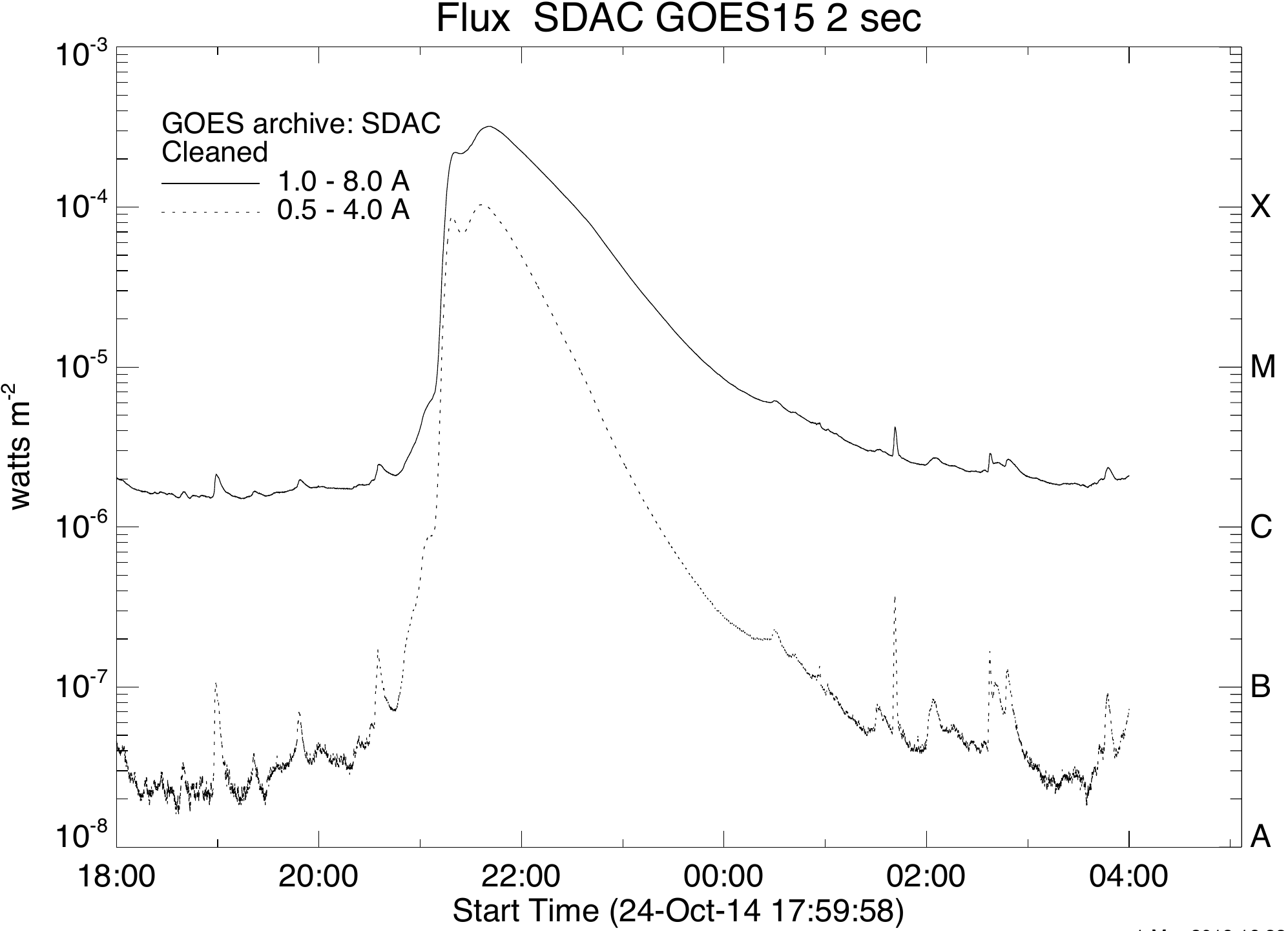}
    \caption{}
  \end{subfigure}
  \begin{subfigure}[]{0.75\textwidth}
    \centering
    \includegraphics[width=1\linewidth]{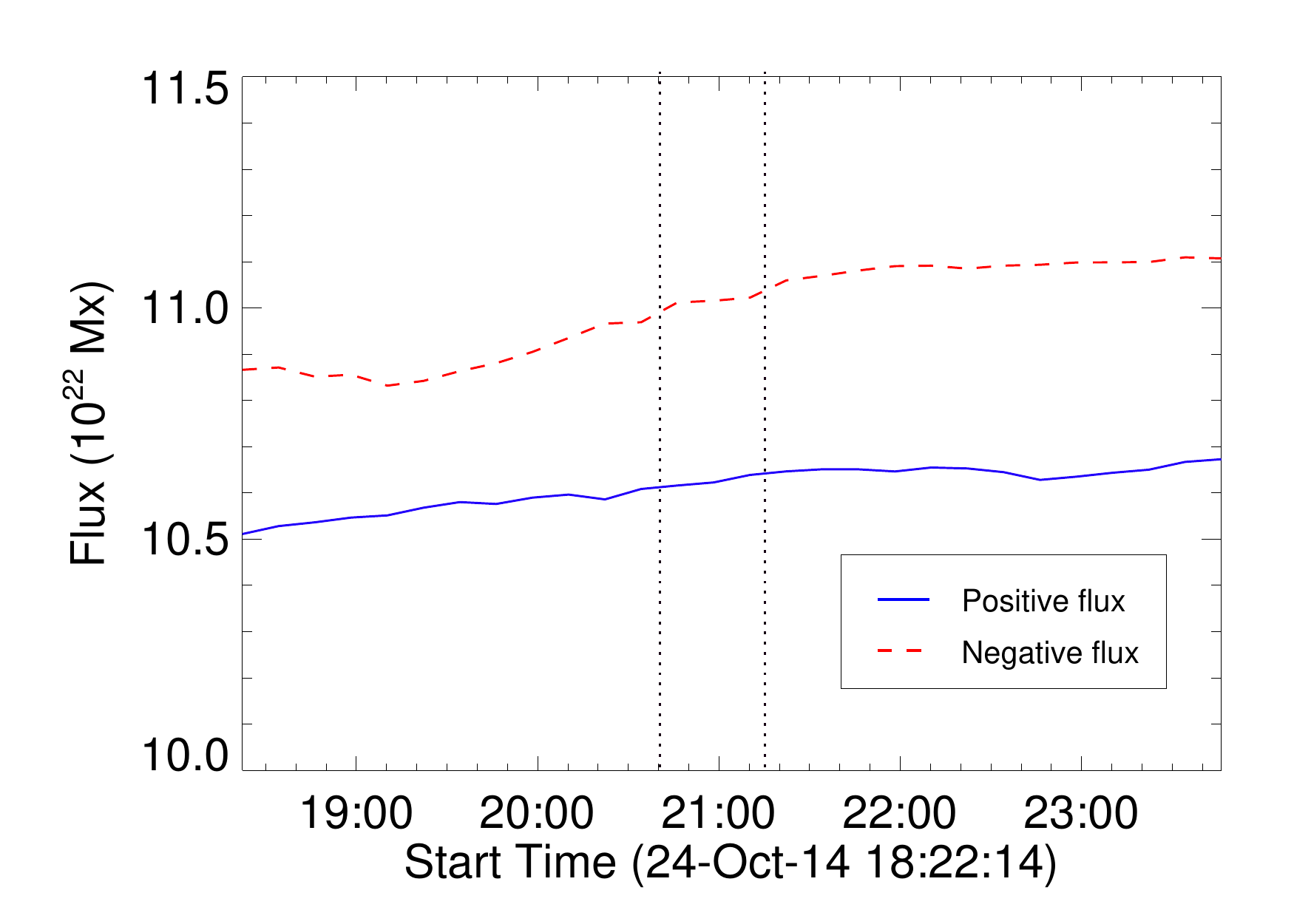}
    \caption{}
  \end{subfigure}
    \caption{(a) GOES 15 X-ray flux for AR 12192 on 24th October, 2014 plotted with time during the X3.1 flare event. Notable is the peak around 21:15 UT  indicating the flare. (b) The evolution of positive and negative magnetic flux at the photospheric boundary during the flare. The vertical dashed lines mark the interval between onset and peak of the flare. Importantly, there is no appreciable flux change within the interval. }
  \label{f:obs}
\end{figure}

The confined X3.1 flare was of long duration, lasting for 6 to 7 hours as shown in  Figure \ref{f:obs}(a). The figure shows the GOES 15 X-ray flux observed during this event in the 1-8 \AA~ and 0.5-4 \AA~ channels. It should be noted that no appreciable change in the vertical magnetic field flux was recorded during this period at the photospheric boundary. This is shown in Figure \ref{f:obs}(b) which  depicts the evolution of negative (dashed line in red)  and positive magnetic fluxes (continuous line in blue), calculated by using the photospheric vector magnetograms from the Heliospheric Magnetic Imager (HMI)\citep{2012SoPh..275..229S} on board the Solar Dynamics Observatory (SDO)\citep{2012SoPh..275....3P}. The magnetograms are taken from the `hmi.sharp\_cea\_720s data series' that provides full-disk vector magnetograms of the Sun with a temporal cadence of 12 minutes and a spatial resolution of $0''.5$. In order to obtain the magnetic field on a Cartesian grid, the magnetogram  is initially remapped on to a Lambert cylindrical equal-area (CEA) projection and then transformed into the heliographic coordinates \citep{1990SoPh..126...21G}.  The  dotted vertical lines mark the beginning and peak phase of the flare.  Hence, to a good approximation, the vertical magnetic field $B_z$ at the bottom boundary remains constant during the interval. Accordingly, the photosphere can be approximated to be line tied --- a boundary condition used in the simulations discussed later in the paper.

\begin{figure}[hp]
  \centering
  \begin{subfigure}[]{0.85\textwidth}
    \centering
    \includegraphics[width=1\linewidth]{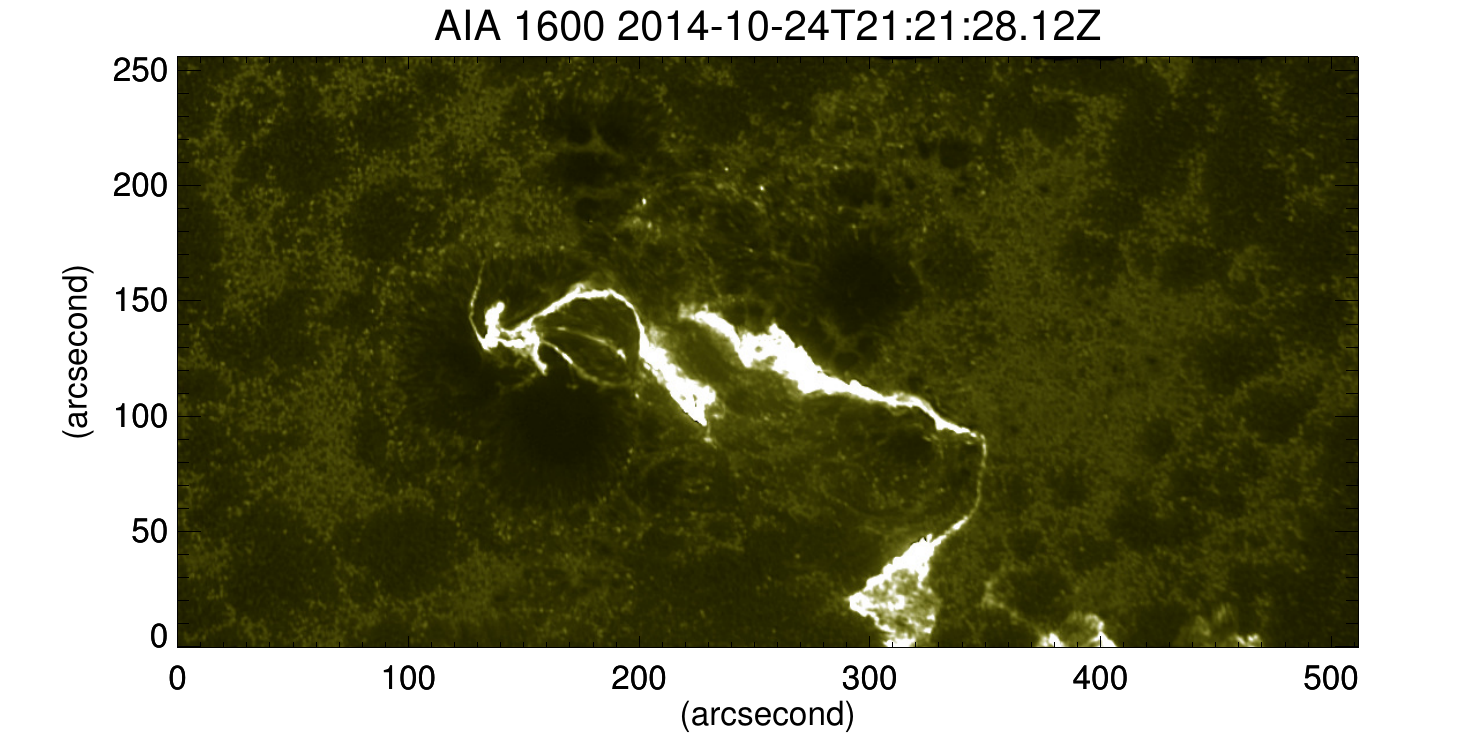}
    \caption{}
  \end{subfigure}
  \begin{subfigure}[]{0.85\textwidth}
    \centering
    \includegraphics[width=1\linewidth]{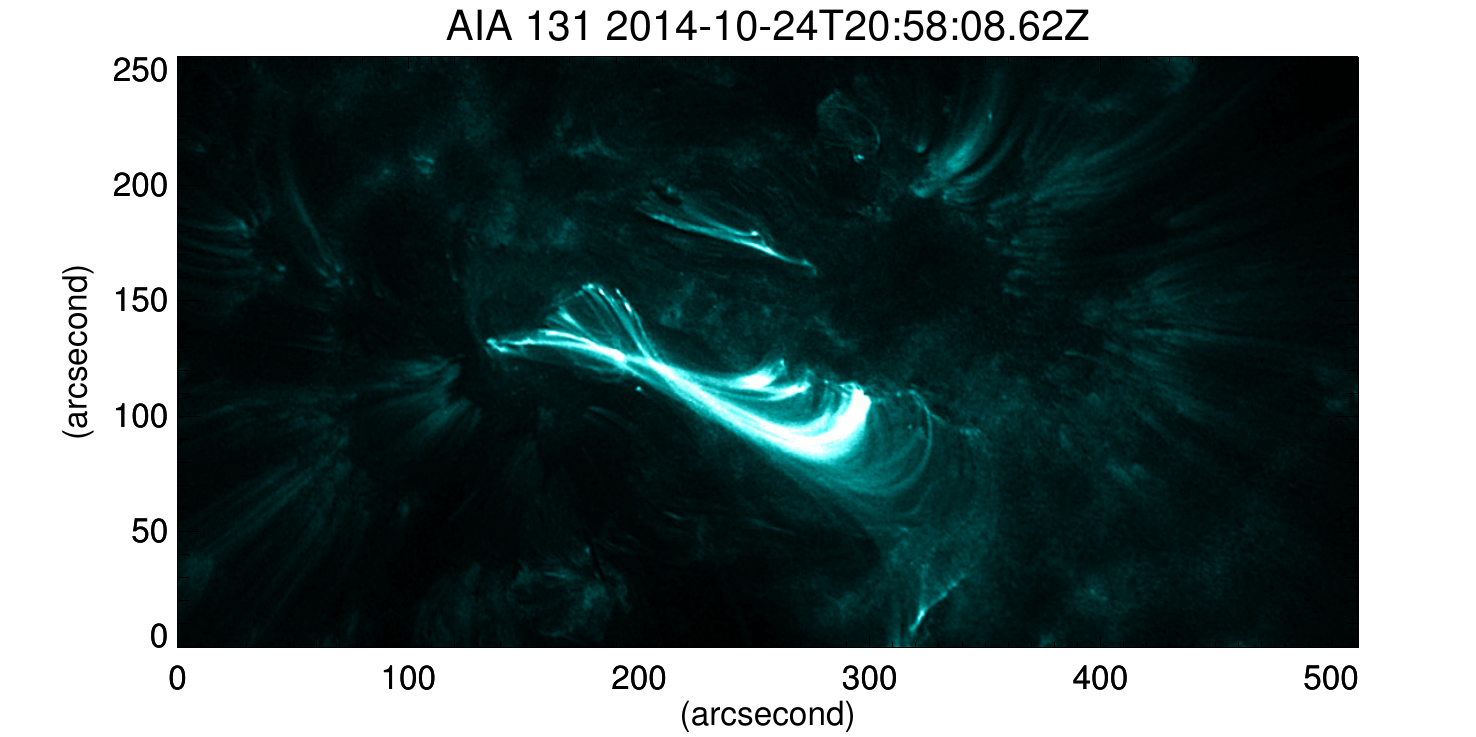}
    \caption{}
  \end{subfigure}
   \caption{The AR 12192 observed in AIA 1600 \AA~ (panel a) during the flare at 24th October, 2014 at 21:21 UT and AIA 131 \AA~ (panel b) at 20:58 UT. The abscissa and ordinate are in arcsecond with one unit corresponding to a physical length of 720 km. Important is the circular brightening located approximately between 150 and 200 arcsecond along the abscissa. }
  \label{f:aia}
\end{figure}

Importantly, a circular brightening was observed in the chromospheric flare ribbons at the ultra-violet (UV) 1600 \AA~ channel preceded by a brightening of the flaring loops in the extreme-ultra-violet (EUV) channel 131 \AA~ of the Atmospheric Imaging Assembly (AIA) on board SDO\citep{2012SoPh..275...17L}). The brightenings occur  in the interval  21:20 to 21:35 UT in the 1600 \AA~ channel (Figure \ref{f:aia} (a)) and is co-located with the brightening in the 131 \AA~ channel as seen around 20:58 UT, which is just before the X-class flare. The circular flare ribbons are known to map MFLs constituting the fan plane of a 3D null on the photosphere \citep{2009ApJ...700..559M}. To our knowledge, the generation of the circular ribbon was not reported in the earlier works and is the main focus of the paper. 

\begin{figure}[hp]
  \centering
  \begin{subfigure}[]{0.85\textwidth}
    \centering
    \includegraphics[width=1\linewidth]{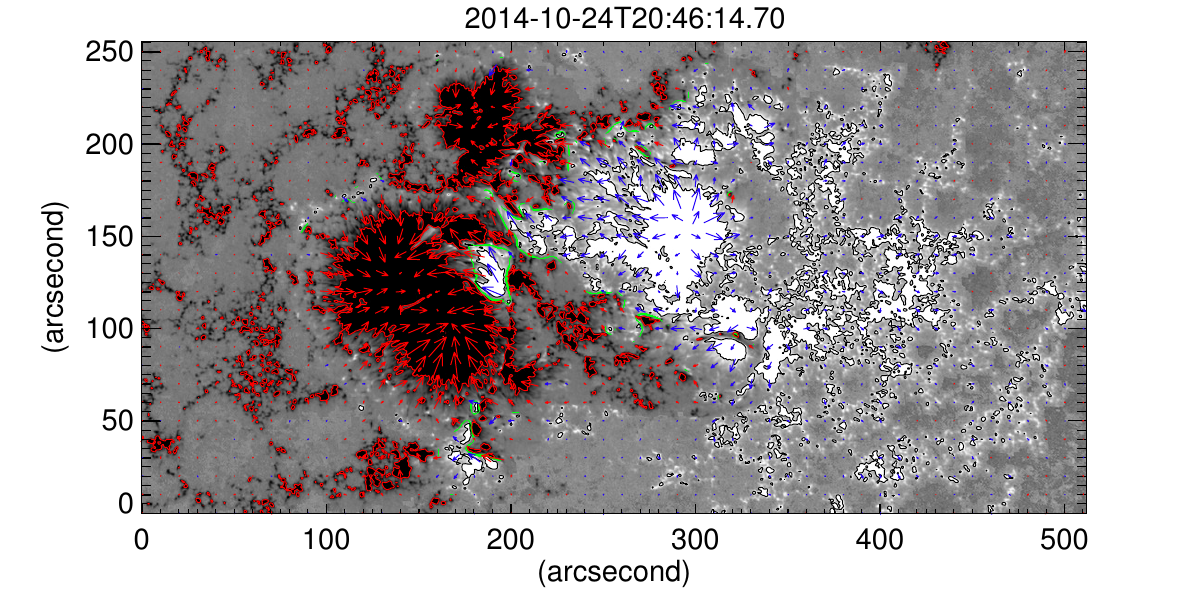}
    \caption{}
  \end{subfigure}
  \begin{subfigure}[]{0.85\textwidth}
    \centering
    \includegraphics[width=1\linewidth]{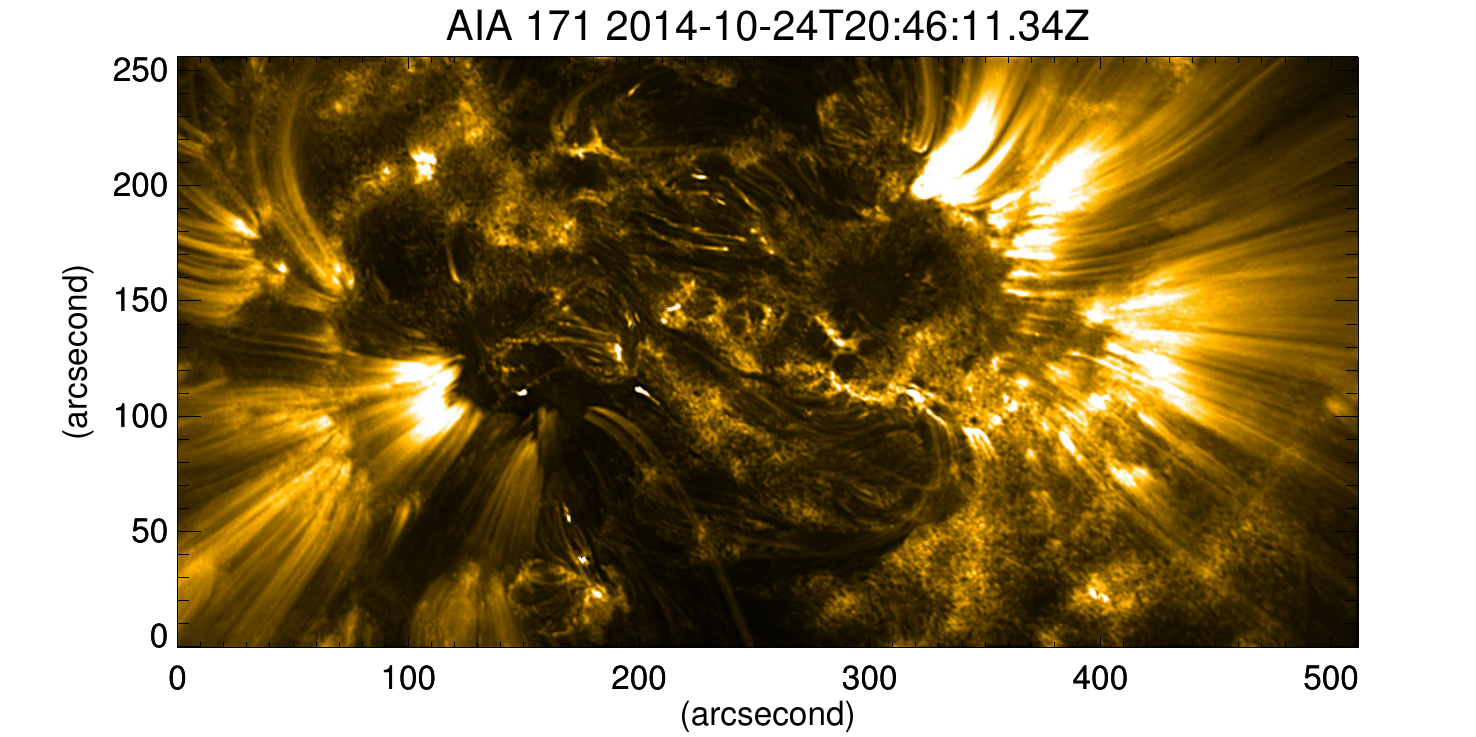}
    \caption{}
  \end{subfigure}
   \caption{(a) Photospheric vector magnetogram from HMI of AR 12192 remapped on a CEA projection at 20:46 UT on 2014-10-24, highlighting the magnetic field line topology before the flare. The black and white contours represent the negative and positive polarities of $B_z$ whereas the red and blue arrows are the vector plots of the transverse magnetic field. (b) AIA 171~\AA~ EUV image of the AR, highlighting the magnetic field line topology before the flare.}
  \label{f:sdo}
\end{figure}

To simulate evolution of such MFLs, we select the vector magnetogram at 20:46 UT, roughly 30 minutes prior to the flare. The Figure \ref{f:sdo}(a) shows the magnetogram of the active region where the positive and the negative polarities of the longitudinal component of the magnetic field are depicted in white and black, and the gray represents the background. The transverse components of the positive and negative fields are shown by blue and red arrows respectively. The PIL is represented in the figure by green lines. The AR is visibly complex, with two main polarities and multiple small-scale features. The MFL topology can be inferred using the extreme ultra-violet (EUV) channel data  as observed in the 171~\AA~, shown in Figure \ref{f:sdo}(b), which plotted on the same CEA spatial grid as in Figure \ref{f:sdo}(a). The EUV coronal loops near the PIL are markedly sheared and twisted, indicating a high degree of complexity in the initial magnetic field topology.

\section{Non-force-free extrapolation of magnetic field}
\label{sec:3}
\subsection{Description of the numerical extrapolation algorithm}
The coronal magnetic field of the AR 12192 is obtained by using the numerical non-force free extrapolation code developed by 
\citet{2008SoPh..247...87H,2008ApJ...679..848H,2010JASTP..72..219H} where ${\bf{B}}$ is constructed as

\begin{equation}
\mathbf{B} = \mathbf{B_1}+\mathbf{B_2}+\mathbf{B_3}; \quad \nabla\times\mathbf{B_i}=\alpha_i\mathbf{B_i}
\label{e:b123}
\end{equation}
with $\alpha_i$ as constant and $i=1,2,3$; rendering each sub-field ${\bf{B}}_i$ to be LFFF and $\alpha_1\ne\alpha_2\ne\alpha_3$. Further, without loss of generality, $\alpha_2=0$ is selected to make $\mathbf{B_2}$ potential. 
Subsequently, an optimal pair $\alpha=\{\alpha_1 , \alpha_3\}$ is obtained by an iterative trial-and-error method which finds the pair that minimizes the average deviation between the observed ($\mathbf{B}_t$) and the calculated ($\mathbf{b}_t$) transverse field, as indicated by the following metric:
\begin{equation}
E_n =\left(\sum_{i=1}^M |\mathbf{B}_{t,i}-\mathbf{b}_{t,i}|\times|\mathbf{B}_{t,i}|\right)/\left(\sum_{i=1}^M |\mathbf{B}_{t,i}|^2\right)
\label{en}
\end{equation}
where $M=N^2$, represents the total number of grids points on the transverse plane. Here, the grid points are weighted with respect to the strength of the observed transverse field, see \citet{2008SoPh..247...87H,2010JASTP..72..219H} for further details.




\subsection{Initial extrapolated NFFF for AR 12192}

\begin{figure}[hp]
  \centering
  \begin{subfigure}[]{0.8\textwidth}
    \centering
    \includegraphics[width=1\linewidth]{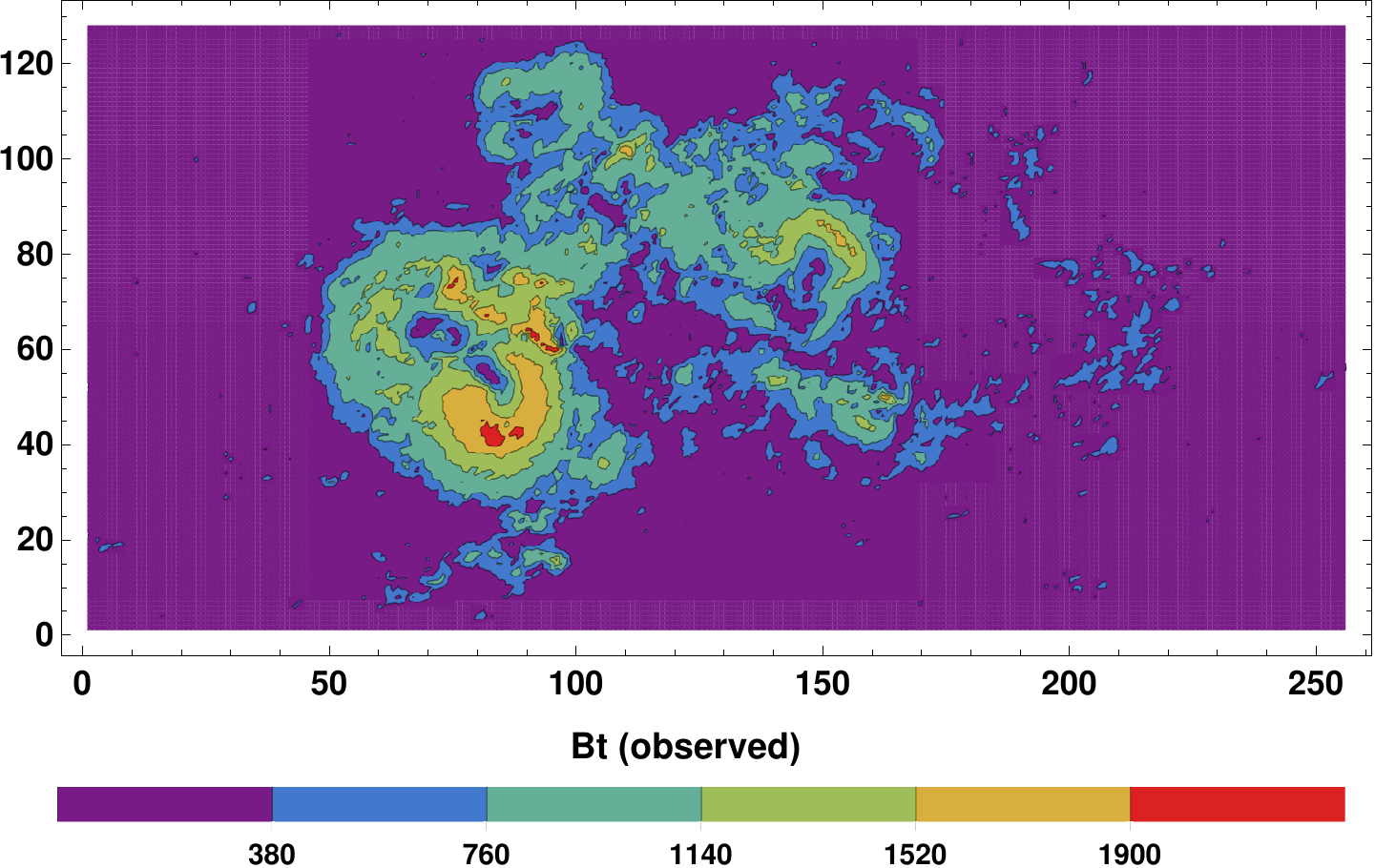}
    \caption{}
  \end{subfigure}
\quad
  \begin{subfigure}[]{0.8\textwidth}
    \centering
    \includegraphics[width=1\linewidth]{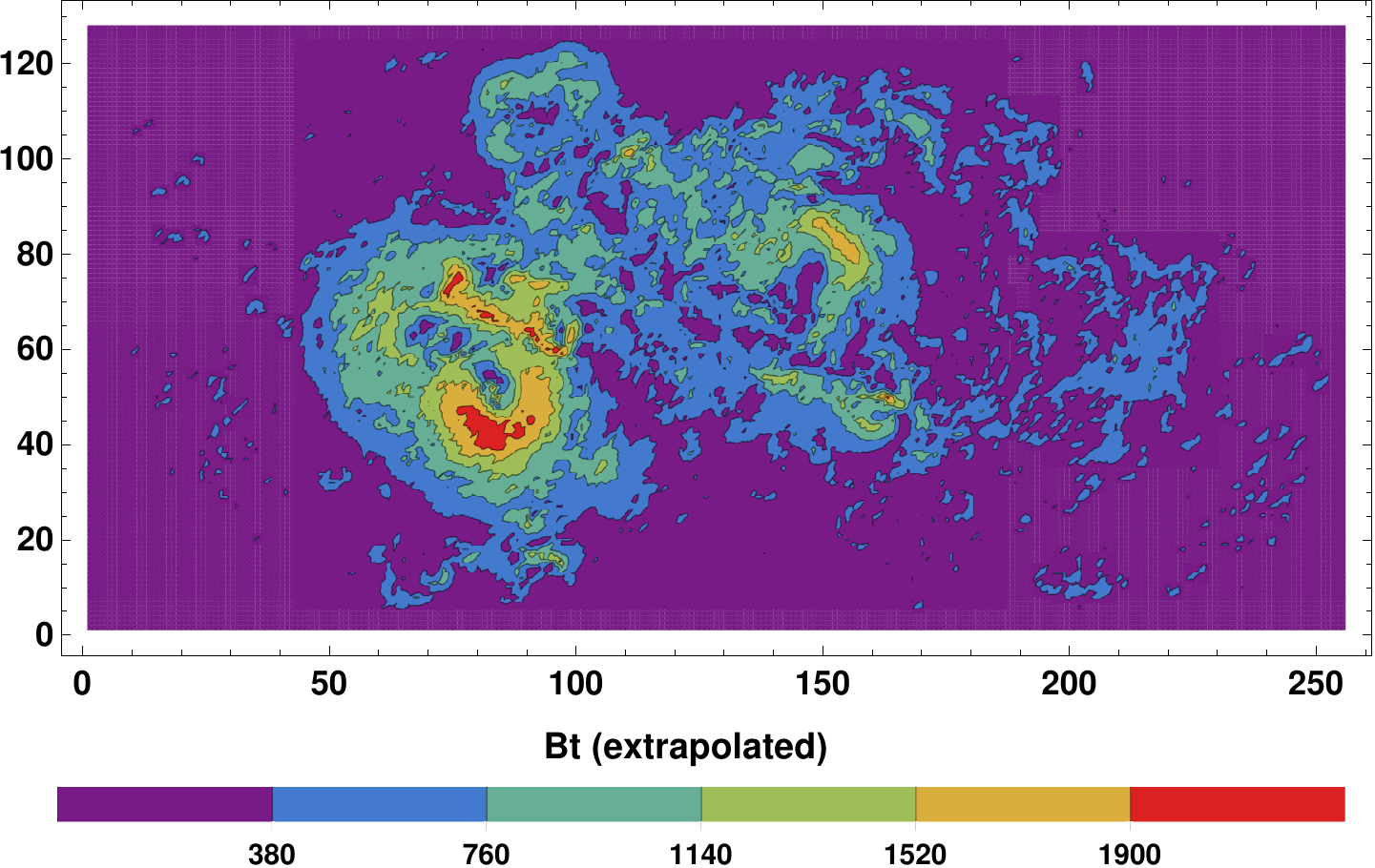}
    \caption{}
  \end{subfigure}
    \caption{Contour plots of the transverse field of the observed (panel a) and extrapolated (panel b) magnetic field shown at the photospheric boundary.}
  \label{f:diff}
\end{figure}
We consider the magnetogram on October 24, 20:46 UT obtained from SDO/HMI. The vector field shown in Figure \ref{f:sdo}(a) corresponds to an original cutout of dimension  $1024\times 512$ pixels. To reduce the computation cost, the field is rescaled and extrapolated over a computational domain having $256\times 128 \times 128$ grids in the $x$, $y$ and $z$ directions.  The corresponding physical extents are 360 Mm in the $x$ direction and 180 Mm in the $y$ and $z$ direction. The best-fit values obtained for the $\alpha$ parameters in this case are $\alpha=\{0.1145,-0.0016\}$ which corresponds to an $E_n=0.31$ (c.f. Equation \eqref{en}). 

\begin{figure}[h!]
  \centering
    \includegraphics[width=0.7\linewidth]{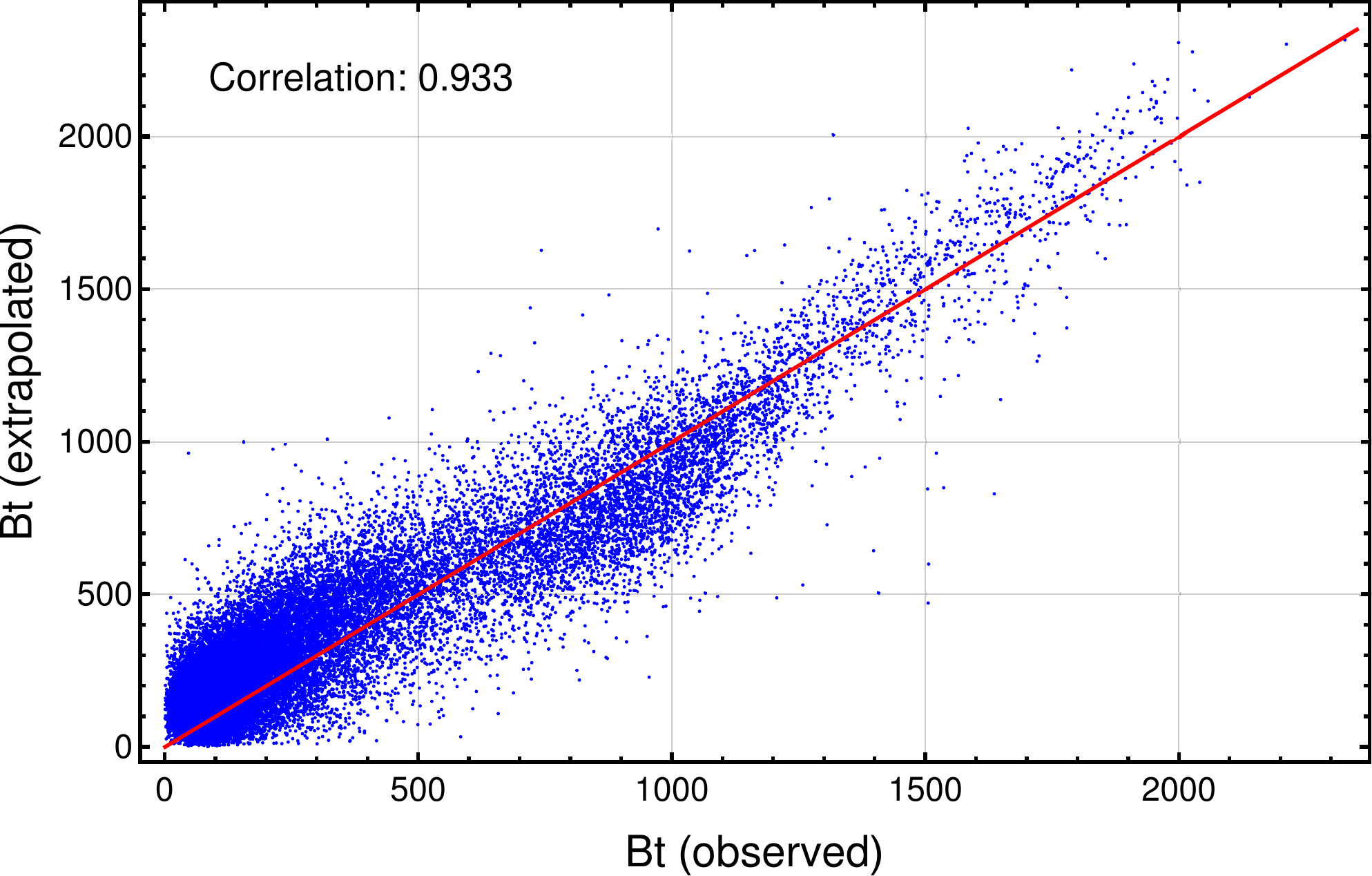}
    \caption{Scatter plot showing the correlation between the observed and extrapolated magnetic field. The red line is the expected profile for perfect correlation.}
  \label{f:scatter}
\end{figure}

The contour plots for the transverse components of the observed and extrapolated fields at the photospheric boundary are shown in Figure \ref{f:diff}. The figure indicates  most of the large scale 
magnetic features  to be well-captured by the extrapolated field. The scatter plot of the observed and the extrapolated fields is shown in Figure \ref{f:scatter}. 
With the perfect correlation---exact agreement of the extrapolated field with the observed one---being marked by the red line, 
the plot documents the agreement to be better in the higher field side. The Pearson-r correlation between the two fields is 0.933, which is acceptable. 

\begin{figure}[hp]
  \centering
  \begin{subfigure}[]{0.75\textwidth}
    \centering
    \includegraphics[width=1\linewidth]{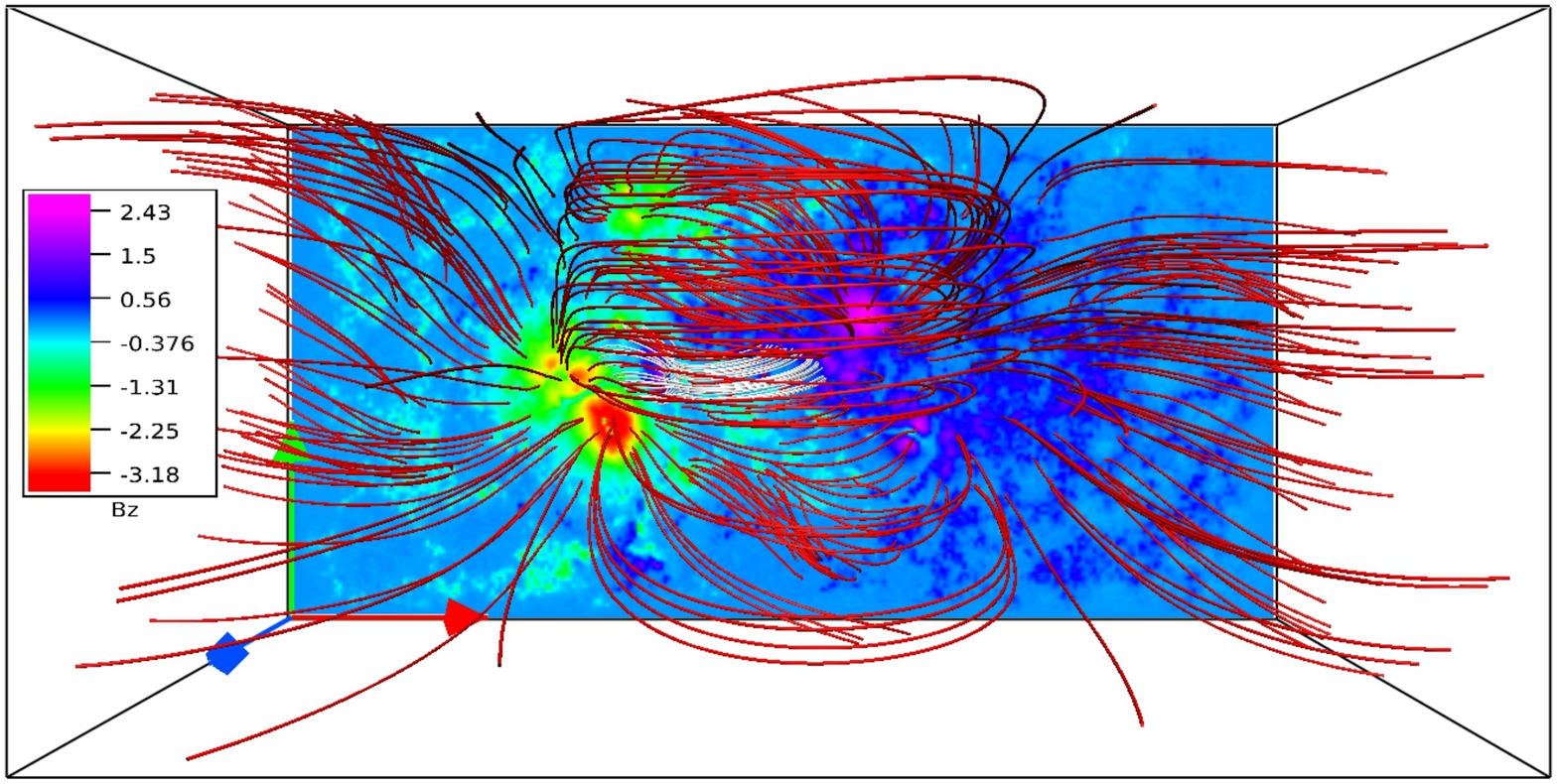}
    \caption{}
  \end{subfigure}
\quad
  \begin{subfigure}[]{0.75\textwidth}
    \centering
    \includegraphics[width=1\linewidth]{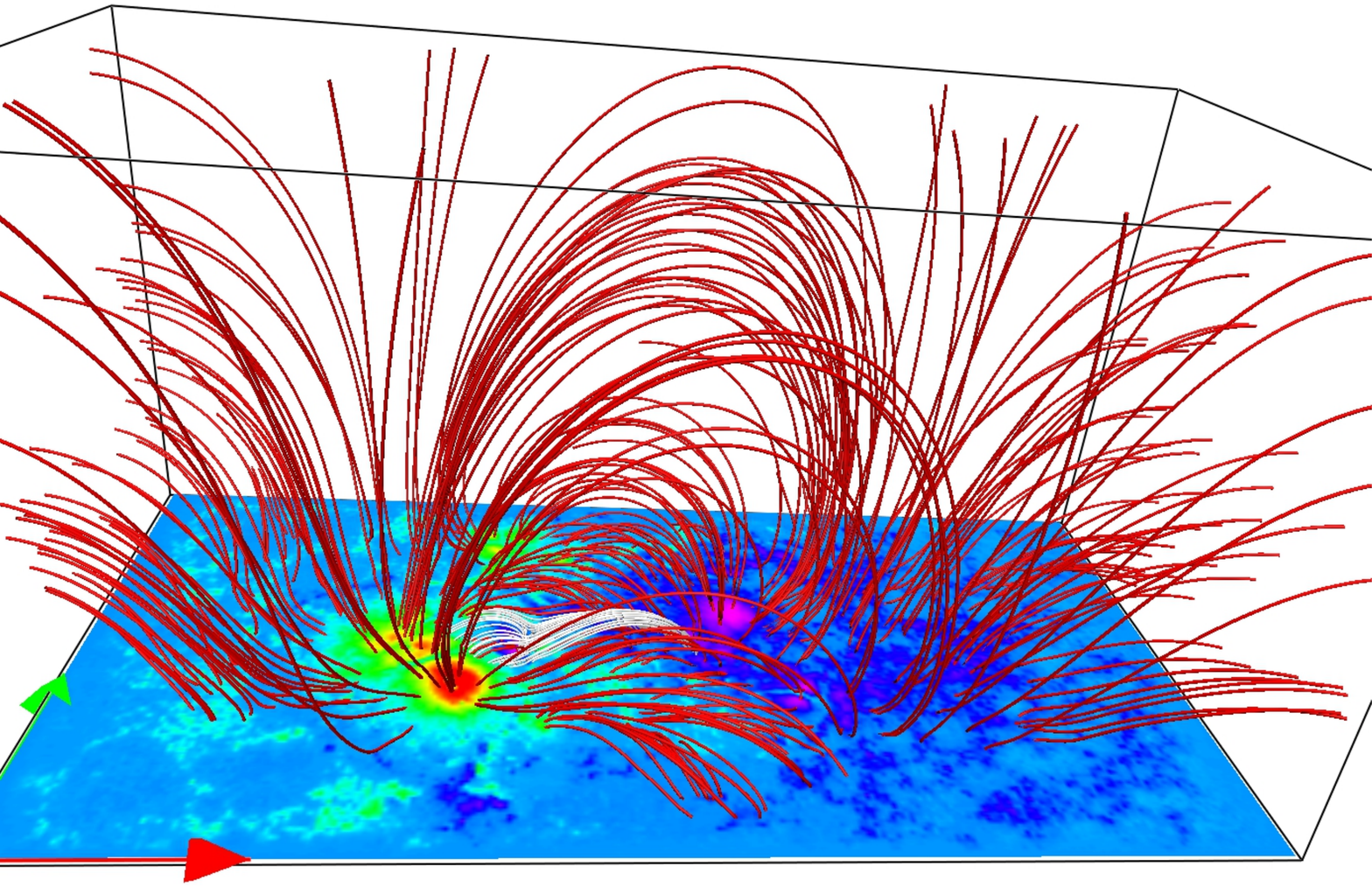}
    \caption{}
  \end{subfigure}
    \caption{Top  (panel a) and side-view (panel b) of the overall NFFF extrapolated magnetic field line topology for the AR 12192 on 2014, October 24, 20:46 UT.  The bottom boundary represents the strength (in kG) of the $B_z$ component of the magnetic field and the extrapolated fields lines are depicted in red and a small set of field lines close to the location of the 3D null---identified in the Figure \ref{f:3dnull}---are depicted in white. }
  \label{f:exp}
\end{figure}

\begin{figure}[hp]
  \centering
  \begin{subfigure}[]{0.7\textwidth}
    \centering
    \includegraphics[width=1\linewidth]{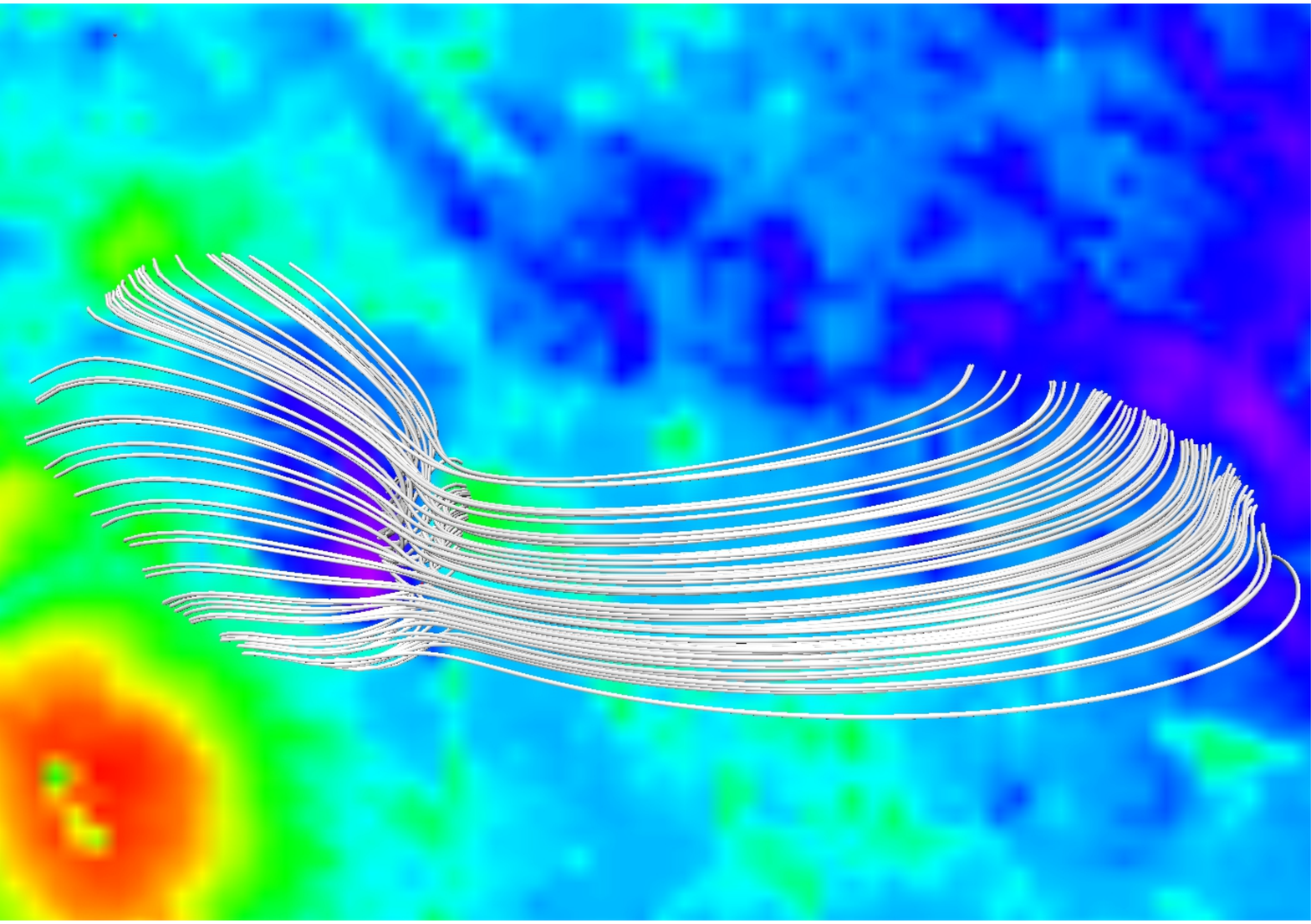}
    \caption{}
  \end{subfigure}
\quad
  \begin{subfigure}[]{0.7\textwidth}
    \centering
    \includegraphics[width=1\linewidth]{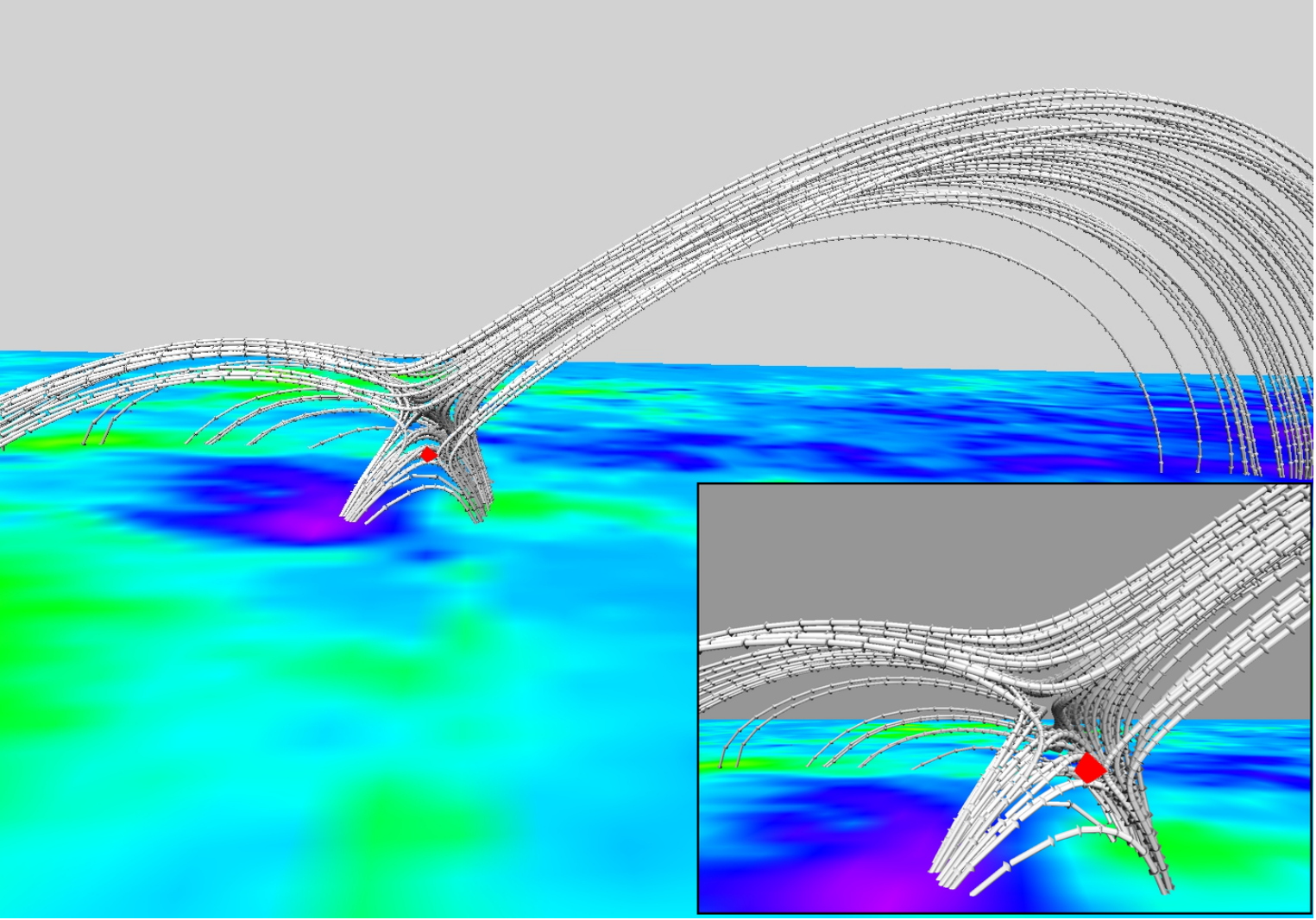}
    \caption{}
  \end{subfigure}
  \caption{The top- (panel a) and side-view (panel b) of the magnetic field lines drawn near one of the polarity-inversion lines, where the flare was later observed. The field line topology indicates the presence of a 3D null, complete with a dome-shaped fan and elongated spine. The red surface inside the field lines represents an isosurface having 2.5\% of the maximal field strength of $B$ and locates the null. The height of the null is roughly 3 Mm from the photosphere. The bottom boundary is same as that of Figure \ref{f:exp}.}
  \label{f:3dnull}
\end{figure}

The top and side views of  MFLs over the full vector magnetogram are shown in Figure \ref{f:exp} with the field lines being printed in red. A smaller set of MFLs in the vicinity of the flaring region (around 21:15 UT) are shown in white. The white MFLs resemble the topology of a 3D magnetic null \citep{1990ApJ...350..672L} and are shown in greater detail in Figure \ref{f:3dnull}. The similarity of MFL morphology of the extrapolated field (panel (b) of Figure \ref{f:exp}) with the observed EUV structure (panel (b) of Figure \ref{f:sdo}) advocates  effectiveness  of the extrapolation. The MFL geometry is characterized by the presence of high and low-lying loops. Notably the low-lying MFLs, depicted in white,  connecting the weak positive polarity 
with the surrounding negative polarity regions generate the 3D null. Figure \ref{f:3dnull}(a) corroborates  the  3D null to be complete with a dome shaped fan and an elongated spine. The panel (b) of Figure \ref{f:3dnull} depicts MFLs on a stack of planes which are approximately tangential to the spine. The MFLs are overlaid with an isosurface (in red) of $|{\bf{B}}|$ having an iso-value which is 2.5\% of its maximum (magnified in the inset). The isosurface locates the 3D null. The height of the null point is roughly 3 Mm from the photospheric plane. Notably, the  MFLs constituting the dome  intersect the bottom boundary to generate footpoints that are distributed in a circular pattern. The MFLs below the null point form an elongated arcade, as seen in the inset of Figure \ref{f:3dnull}.

\begin{figure}[hp]
  \centering
  \begin{subfigure}[]{0.75\textwidth}
    \centering
    \includegraphics[width=1\linewidth]{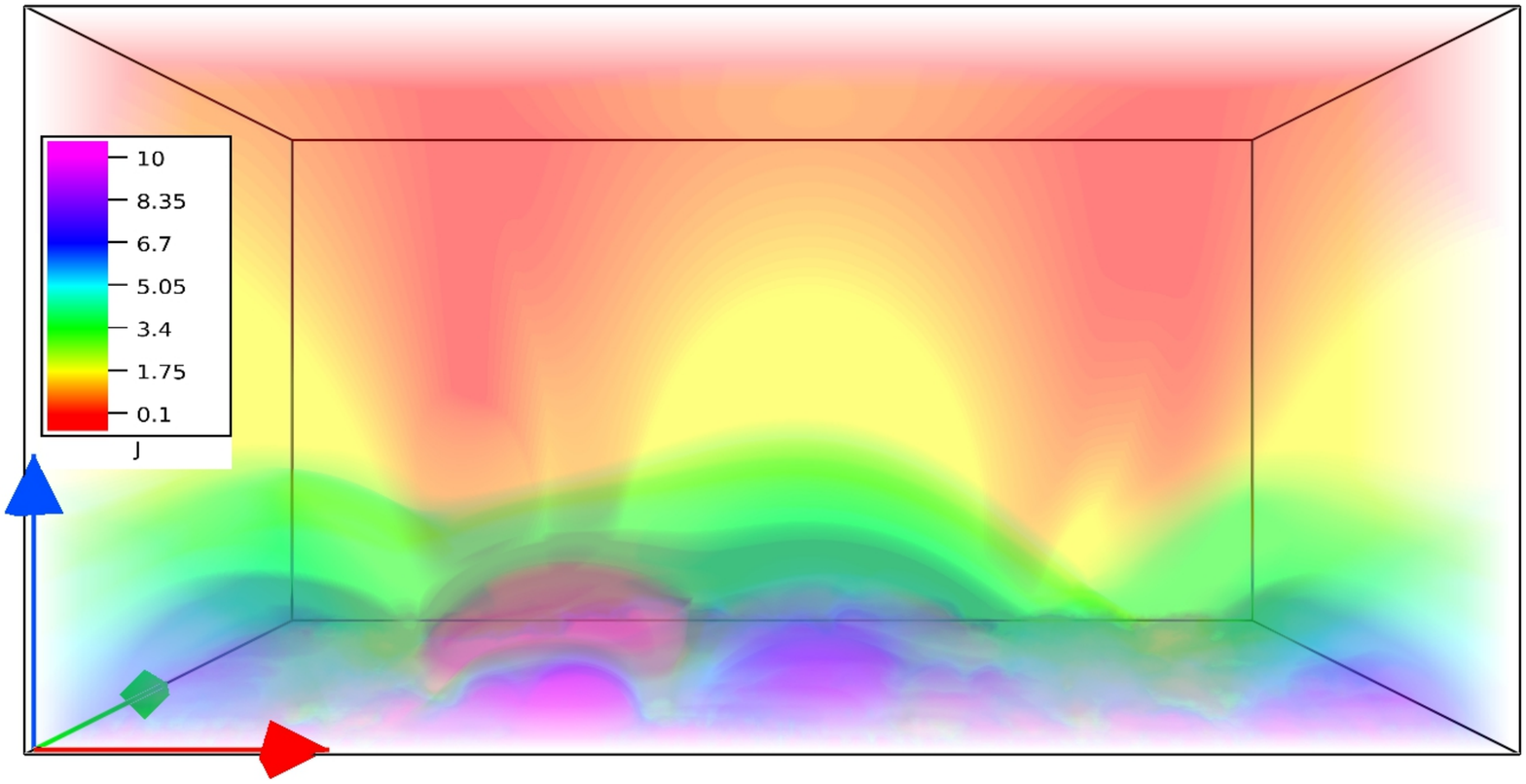}
    \caption{}
  \end{subfigure}
\quad
  \begin{subfigure}[]{0.75\textwidth}
    \centering
    \includegraphics[width=1\linewidth]{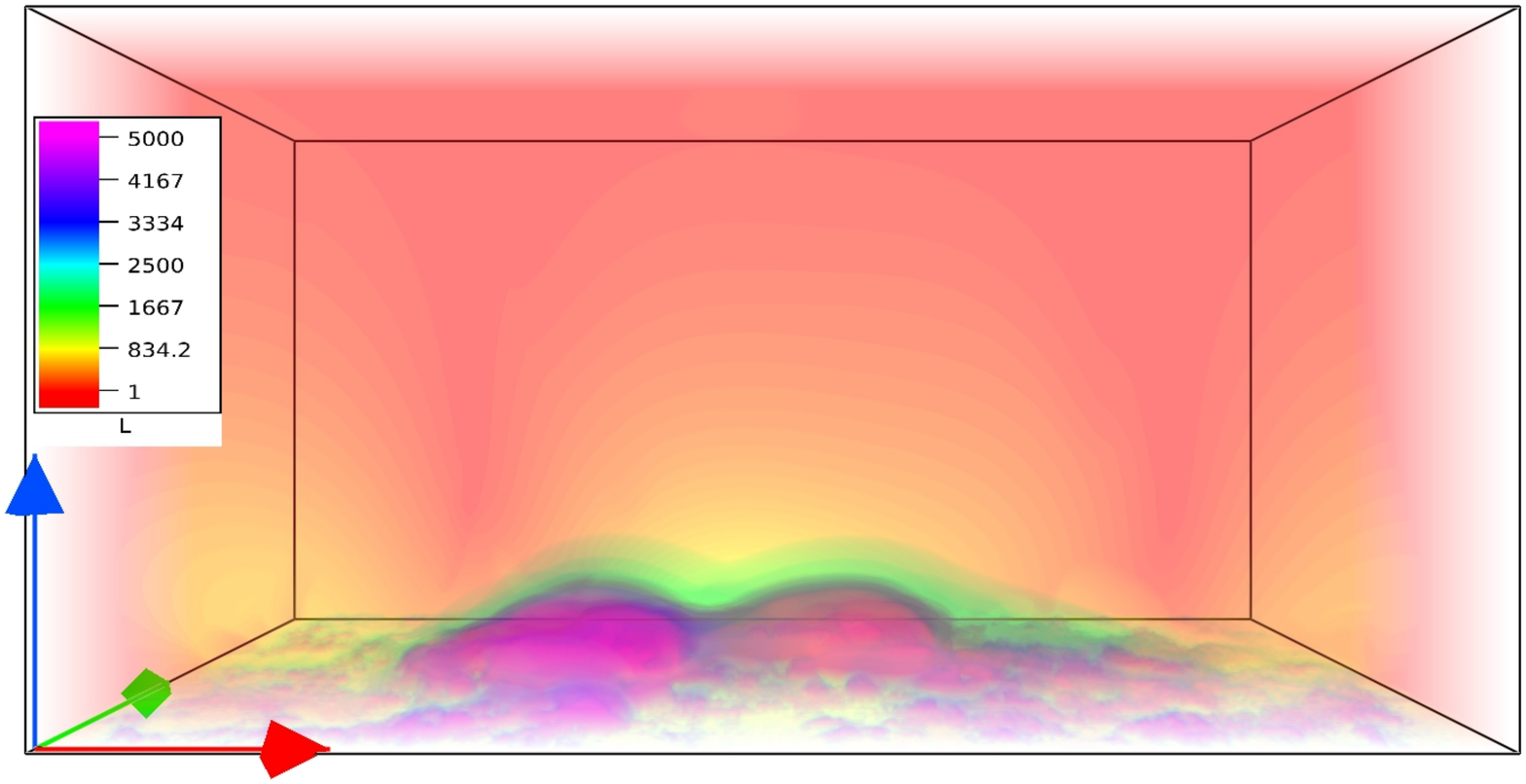}
    \caption{}
  \end{subfigure}
  \caption{The spatial distribution of volume current density (panel a) and Lorentz force (panel b) in arbitrary units. Notably, appreciable current is present throughout the volume while most of the force is present only near the bottom boundary which sharply falls to zero with increase in height.}
  \label{f:clor}
\end{figure}

The direct volume renderings of volume current density $|\mathbf{J}|$ and Lorentz force $|\mathbf{L}|$ are depicted in Figure \ref{f:clor}. Noticeably,  the regions of large Lorentz force and high current overlap with those of high values of $|B_z|$, which can be realized by a direct comparison with Figure \ref{f:exp} (b). The values for $|\mathbf{J}|$ and $|\mathbf{L}|$ are mentioned in arbitrary units as we are mostly interested in their variation with height. The figure reveals a sharp decay of the Lorentz force with height (by a factor of 1/5000) while the current shows a decay by only a factor of 1/100. The current thus becomes more and more field-aligned with increasing height, ultimately making the magnetic field force-free in the asymptotic limit.

\section{Numerical model}
The evolution is governed by the incompressible Navier-Stokes MHD equations under the assumption of thermal homogeneity and perfect electrical 
conductivity  \citep{2010PhPl...17k2901B,2014PhPl...21e2904K,2015PhPl...22h2903K}: 

\begin{subequations}
\begin{align}
\label{stokes}
&  \frac{\partial{\bf{v}}}{\partial t} 
+ \left({\bf{v}}\cdot\nabla \right) {\bf{ v}} =-\nabla p
+\left(\nabla\times{\bf{B}}\right) \times{\bf{B}}+\frac{\tau_a}{\tau_\nu}\nabla^2{\bf{v}},\\  
\label{incompress1}
&  \nabla\cdot{\bf{v}}=0, \\
\label{induction}
&  \frac{\partial{\bf{B}}}{\partial t}=\nabla\times({\bf{v}}\times{\bf{B}}), \\
\label{solenoid}
 &\nabla\cdot{\bf{B}}=0, 
\end{align}  
\label{e:mhd}
\end{subequations}
written in the usual notations in dimensionless form. The normalizations for various terms in Equation \eqref{e:mhd} are as follows
\begin{equation}
\label{norm}
{\bf{B}}\longrightarrow \frac{{\bf{B}}}{B_0},\quad{\bf{v}}\longrightarrow \frac{\bf{v}}{v_a},\quad
 L \longrightarrow \frac{L}{L_0},\quad t \longrightarrow \frac{t}{\tau_a},\quad
 p  \longrightarrow \frac{p}{\rho {v_a}^2}. 
\end{equation}
The constants $B_0$ and $L_0$ are fixed using the average magnetic field strength and length-scale of the vector magnetogram respectively. Here, $v_a \equiv B_0/\sqrt{4\pi\rho_0}$ is the Alfv\'{e}n speed and $\rho_0$ is the constant mass density. The constants $\tau_a$ and $\tau_\nu$, having dimensions of time, represent the Alfv\'{e}n transit time ($\tau_a=L_0/v_a$) and viscous diffusion time scale ($\tau_\nu= L_0^2/\nu$), respectively. The kinematic viscosity is denoted by $\nu$. The ratio  $\tau_a/\tau_\nu$ represents an effective viscosity of the system which,  along with the other forces, 
influences the dynamics. 

To solve the  MHD Equations (\ref{stokes})-(\ref{solenoid}), we utilize the well established magnetohydrodynamic numerical model EULAG-MHD {\citep{2013JCoPh.236..608S}}, which is an extension of the hydrodynamic model EULAG predominantly used in atmospheric and climate research
\citep{Prusa20081193}. The pressure perturbation, denoted by $p$, about a thermodynamically uniform ambient state satisfies an elliptic boundary value problem, which is generated by imposing the discretized incompressibility constraint (Equation \ref{incompress1}) on the discrete integral form of the momentum equation (Equation \ref{stokes}); cf.\citep{2010PhPl...17k2901B} and the references therein. An identical procedure involving the gradient of an auxiliary potential in the induction equation (Equation \ref{induction}) is employed to keep ${\mathbf{B}}$ solenoidal,  see \citet{2010ApJ...715L.133G} and \citet{2013JCoPh.236..608S} for details.
For the completeness, here we mention  only important features of the EULAG-MHD and refer the readers to \citet{2013JCoPh.236..608S} and references therein for detailed discussions. The model is based on the spatio-temporally second-order accurate non-oscillatory forward-in-time multidimensional positive definite advection transport algorithm, MPDATA \citep{2006IJNMF..50.1123S}. 
Important is the proven dissipative property of the MPDATA which, intermittently and adaptively, regularizes the under-resolved scales by simulating MRs 
and mimicking the action of explicit subgrid-scale turbulence models \citep{2006JTurb...7...15M} in the spirit of
Implicit Large Eddy Simulations (ILES) \citep{grinstein2007implicit}.
Such ILESs performed with the model have already been successfully utilized to simulate magnetic reconnections (MRs) to understand their role in the development of various magnetic structures in the solar corona \citep{2015PhPl...22a2902K,2016ApJ...830...80K,2017ApJ...840...37P}. The simulations presented  continue to rely on the effectiveness of ILES in regularizing the onset of MRs.

\section{Simulation results and discussions}
The simulations are initialized from a motionless state with the initial magnetic field given by the  NFFF extrapolation and the magnetofluid
idealized to be thermally homogeneous and having perfect electrical conductivity.  The flow is generated as the initial Lorentz force pushes 
the plasma. To ensure the net magnetic flux to be zero in the computational domain, all components of volume ${\bf{B}}$ except for $B_z$, are continued to the boundaries for a given time step \citep{2015PhPl...22a2902K}. At the bottom boundary, $B_z$ is kept constant (line-tied boundary) since the change of magnetic flux at the boundary is minimal (see Figure \ref{f:obs}(b)). For the simulation, we  set the dimensionless constant $\tau_a / \tau_\nu \approx 7 \times 10^{-3}$, which is roughly two orders of magnitude larger than its coronal value. The higher value of $\tau_a / \tau_\nu$ speeds up the relaxation because of a more efficient viscous dissipation without affecting magnetic topologies. The density is set to $\rho_0=1$ and kinematic viscosity to $\nu = 0.002$, in scaled units. The spatial unit step $\Delta x = 0.0078$, while the time step 
is taken as $\Delta t = 5\times10^{-3}$ to satisfy the Courant-Friedrichs-Lewy (CFL) stability condition \citep{courant1967partial}. The results presented here pertain to a run for 1000 $\Delta t$ which roughly corresponds to an observation time of one and half hour. Due to the constant mass density, the flow generated in the computation is incompressible---an assumption also used in earlier works \citep{1991ApJ...383..420D,2005A&A...444..961A}. Although the compressibility of the fluid is important for the thermodynamics of coronal loops \citep{2002ApJ...577..475R}, our focus for the present is on their magnetic topology only. Notably, the $R_M$ throughout  the simulation is infinity expect during MRs facilitated by  the MPDATA driven dissipation. 

\begin{figure}[hp]
  \centering
  \begin{subfigure}[]{0.45\textwidth}
    \centering
    \includegraphics[width=1\linewidth]{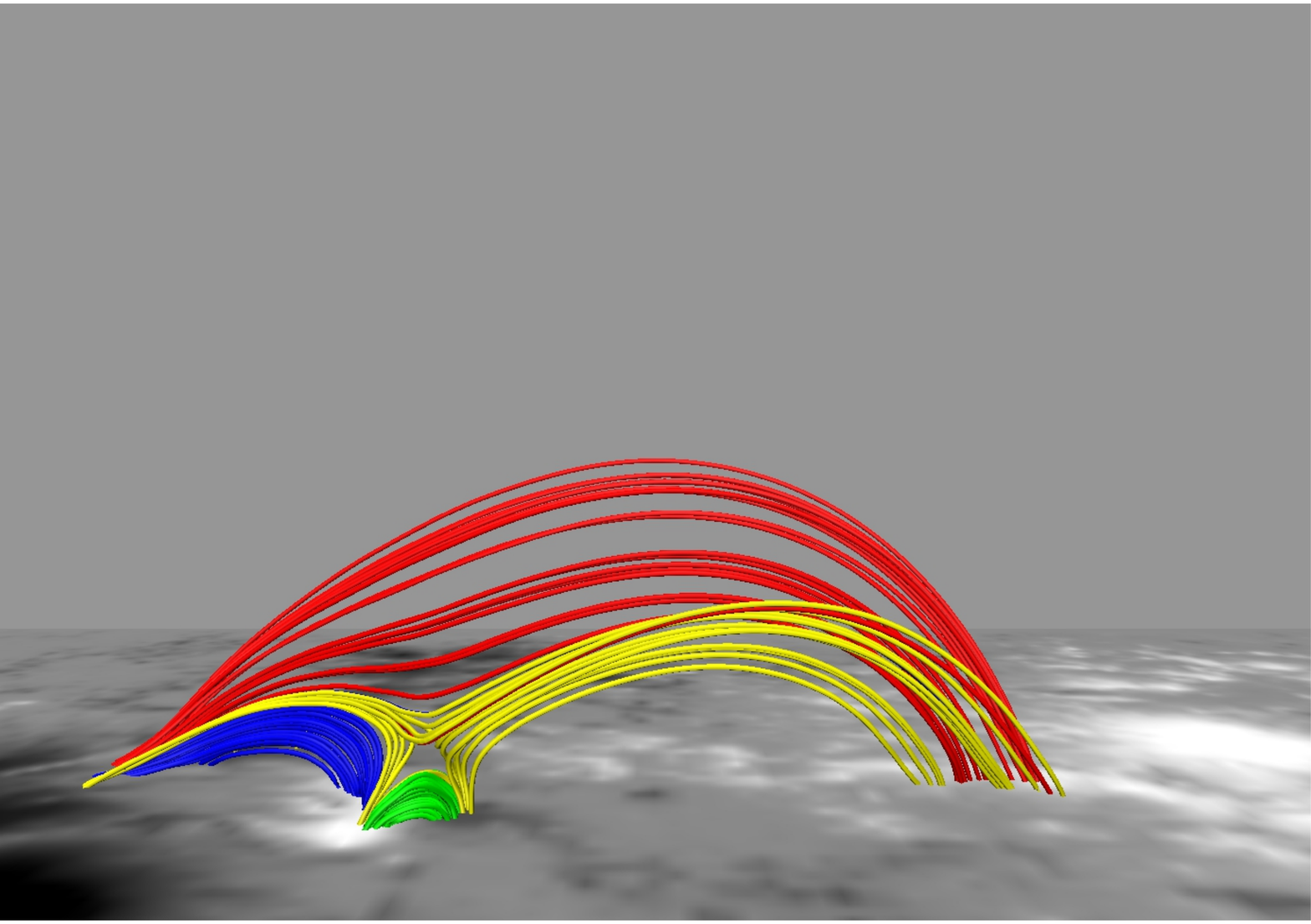}
    \caption{}
  \end{subfigure}
\quad
  \begin{subfigure}[]{0.45\textwidth}
    \centering
    \includegraphics[width=1\linewidth]{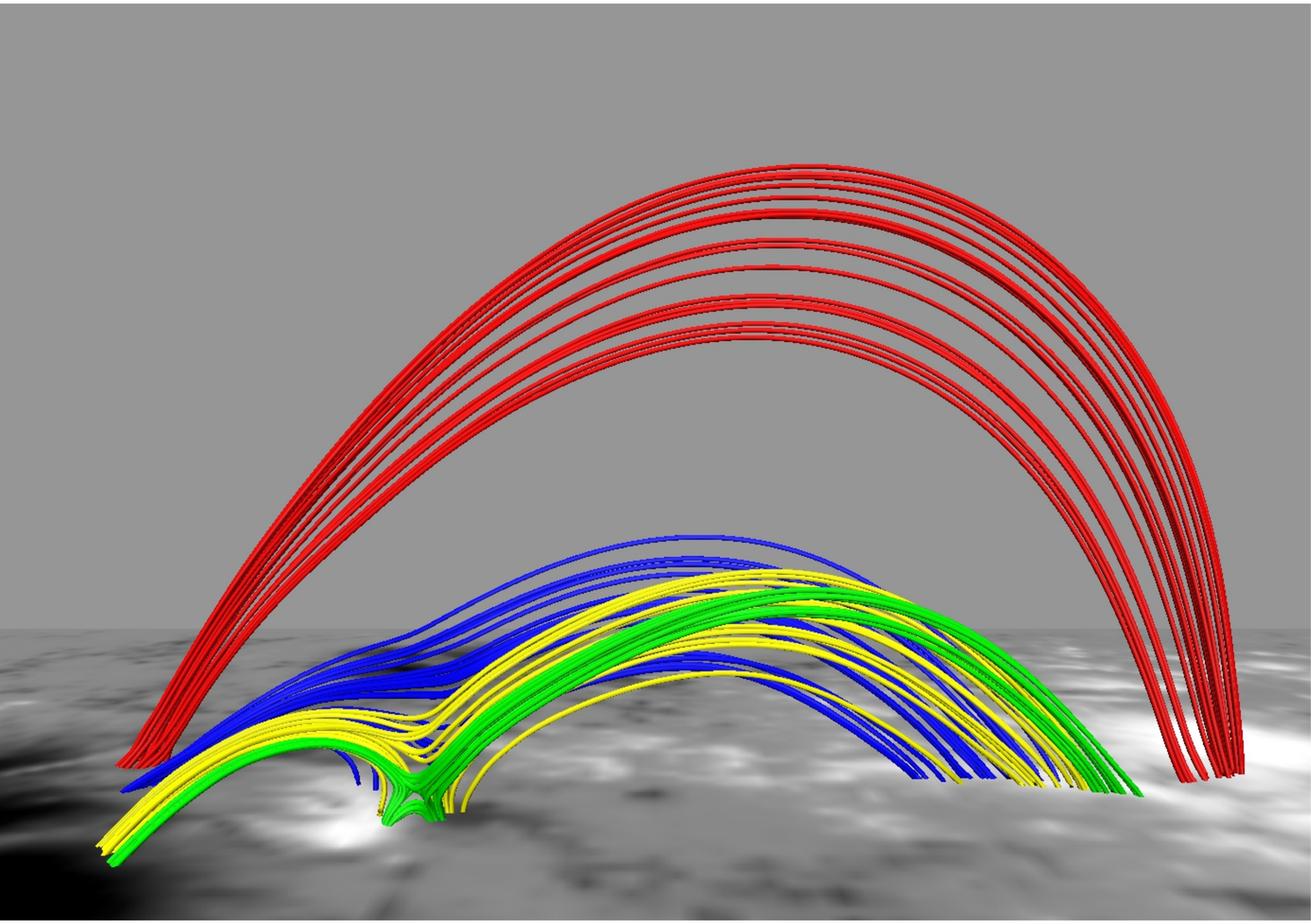}
    \caption{}
  \end{subfigure}
\quad
\begin{subfigure}[]{0.45\textwidth}
    \centering
    \includegraphics[width=1\linewidth]{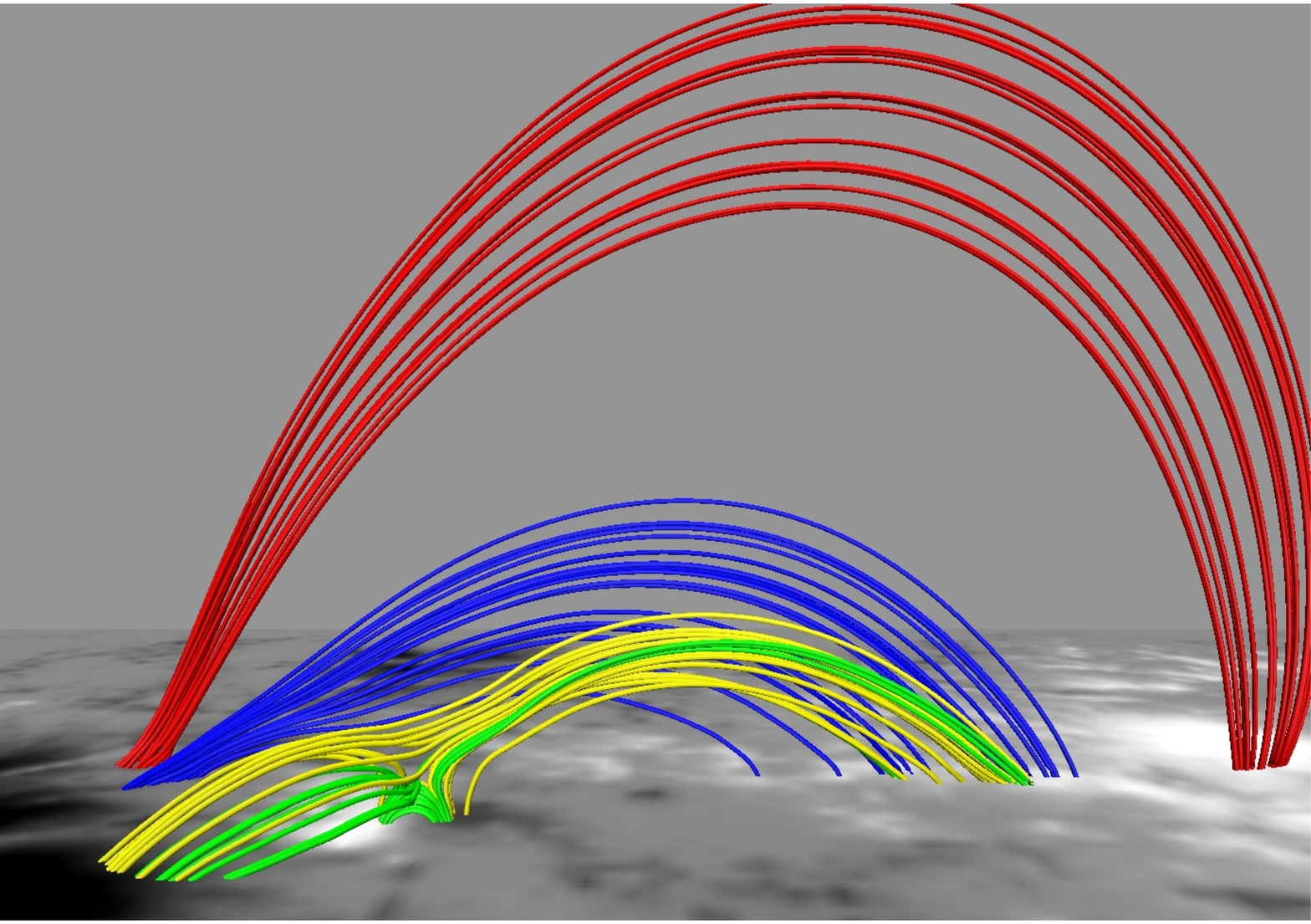}
    \caption{}
  \end{subfigure}
\quad
  \begin{subfigure}[]{0.45\textwidth}
    \centering
    \includegraphics[width=1\linewidth]{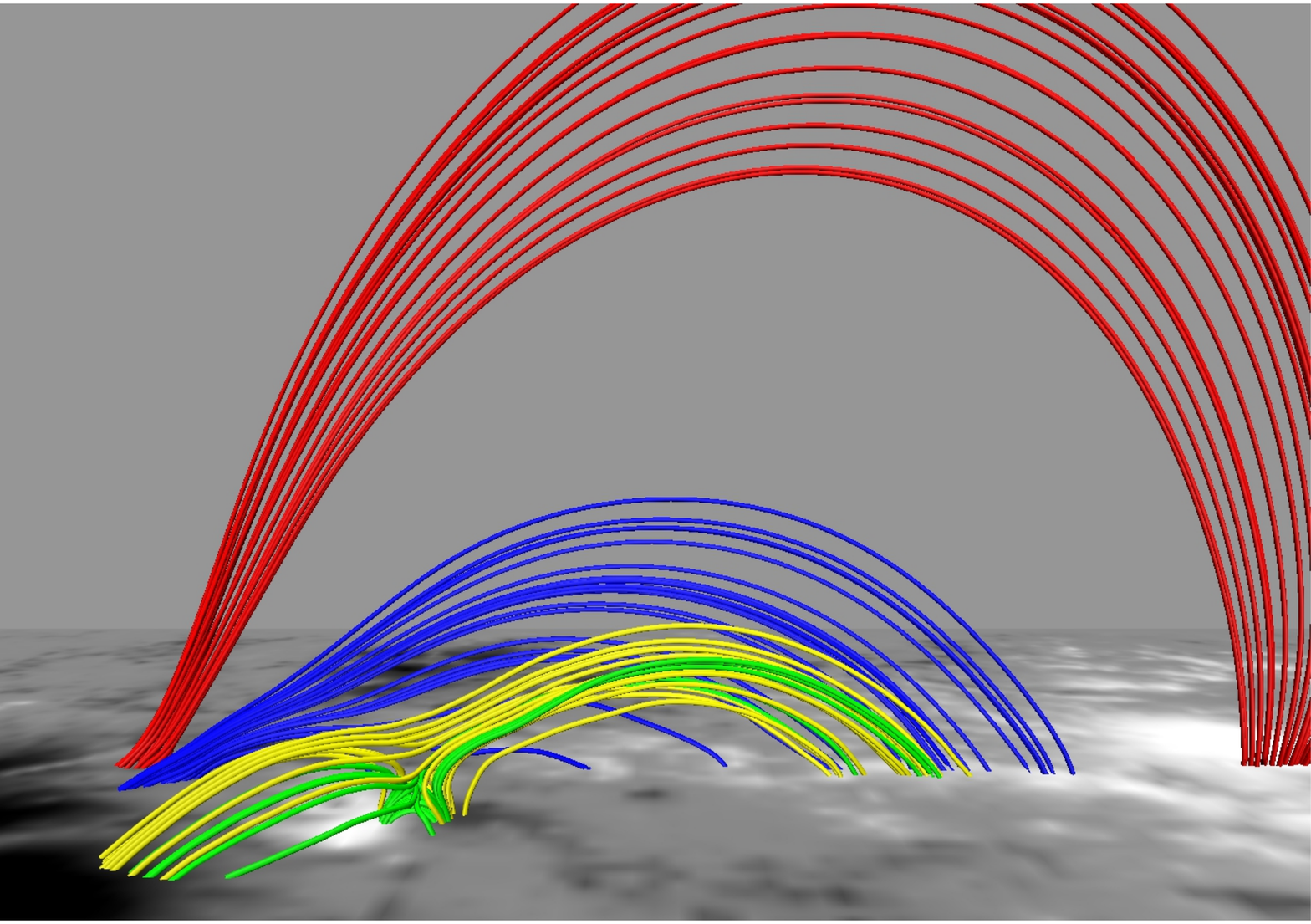}
    \caption{}
  \end{subfigure}
\quad
\begin{subfigure}[]{0.45\textwidth}
    \centering
    \includegraphics[width=1\linewidth]{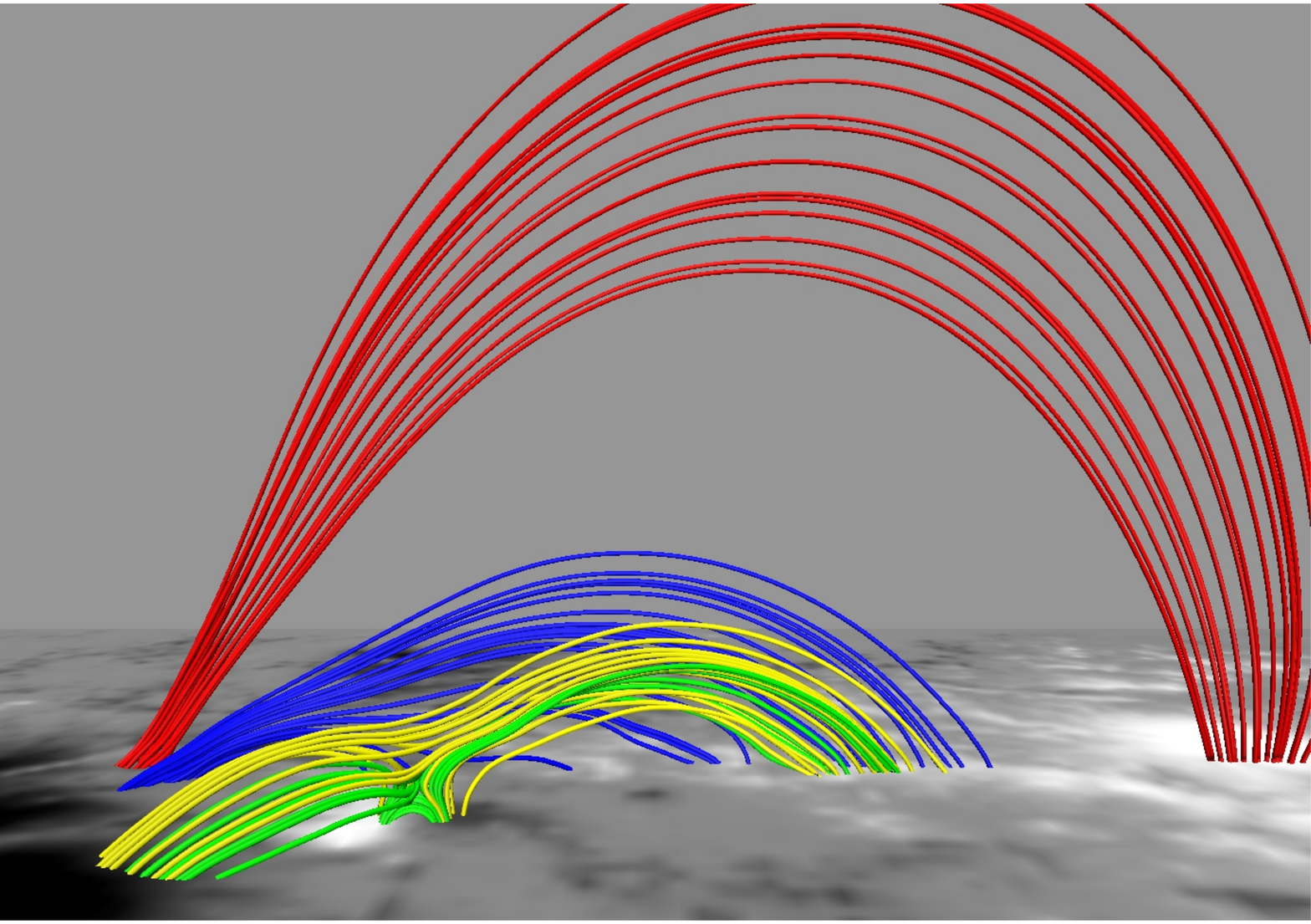}
    \caption{}
  \end{subfigure}
\quad
  \begin{subfigure}[]{0.45\textwidth}
    \centering
    \includegraphics[width=1\linewidth]{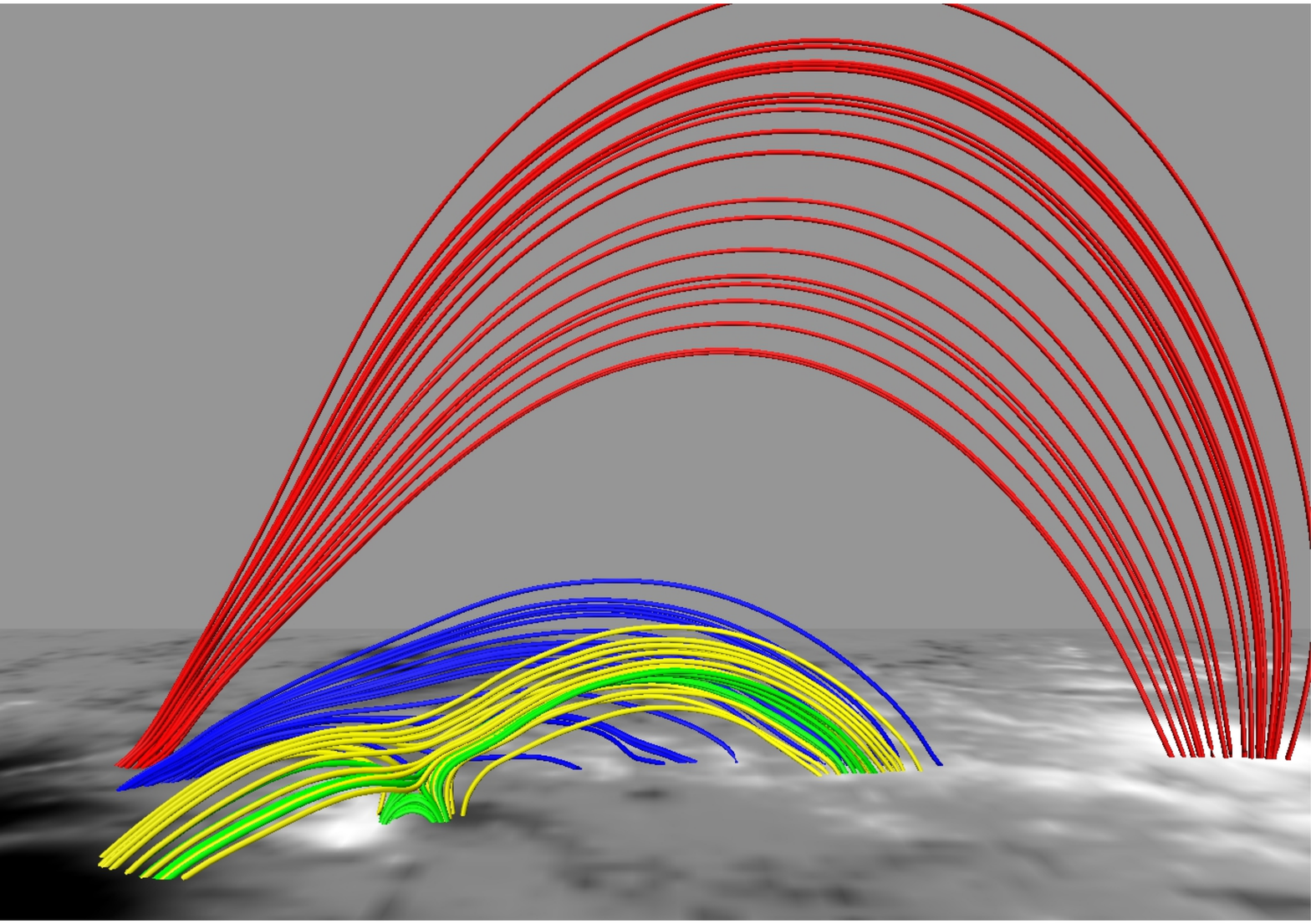}
    \caption{}
  \end{subfigure}
\caption{Side view of evolution of four sets of magnetic field lines close to location of the 3D null, shown at $t$ = 0, 200, 400, 600, 800 and 1000 in panels (a) to (f) respectively. The bottom boundary in all the panels represents the strength $B_z$ on the photospheric plane as in Figure \ref{f:exp} but now in grayscale for clarity. (An animation of this figure is available.)}
 \label{f:recon-side}
\end{figure}
\begin{figure}[hp]
  \centering
  \begin{subfigure}[]{0.45\textwidth}
    \centering
    \includegraphics[width=1\linewidth]{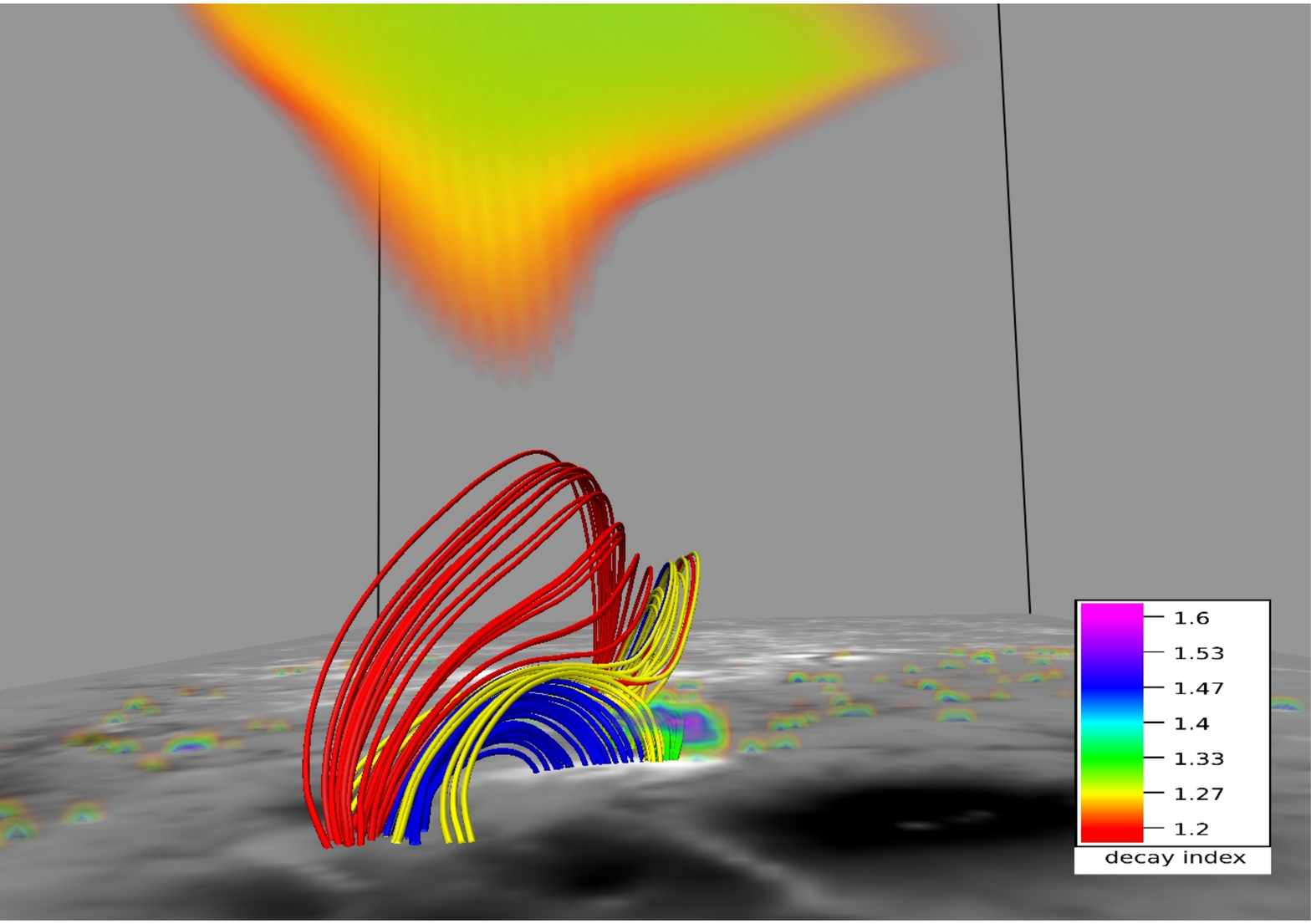}
    \caption{}
  \end{subfigure}
\quad
  \begin{subfigure}[]{0.45\textwidth}
    \centering
    \includegraphics[width=1\linewidth]{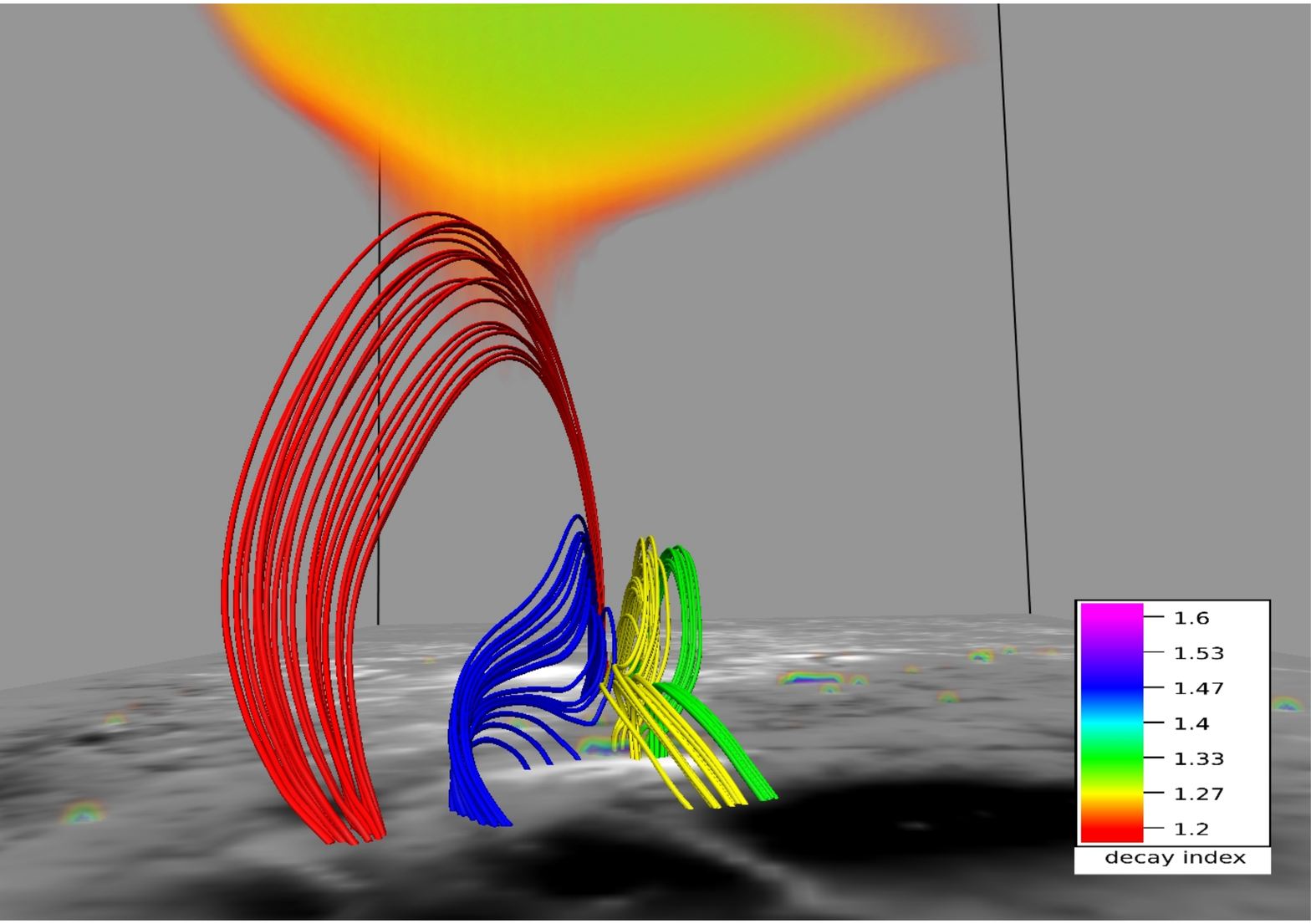}
    \caption{}
  \end{subfigure}
\quad
\begin{subfigure}[]{0.45\textwidth}
    \centering
    \includegraphics[width=1\linewidth]{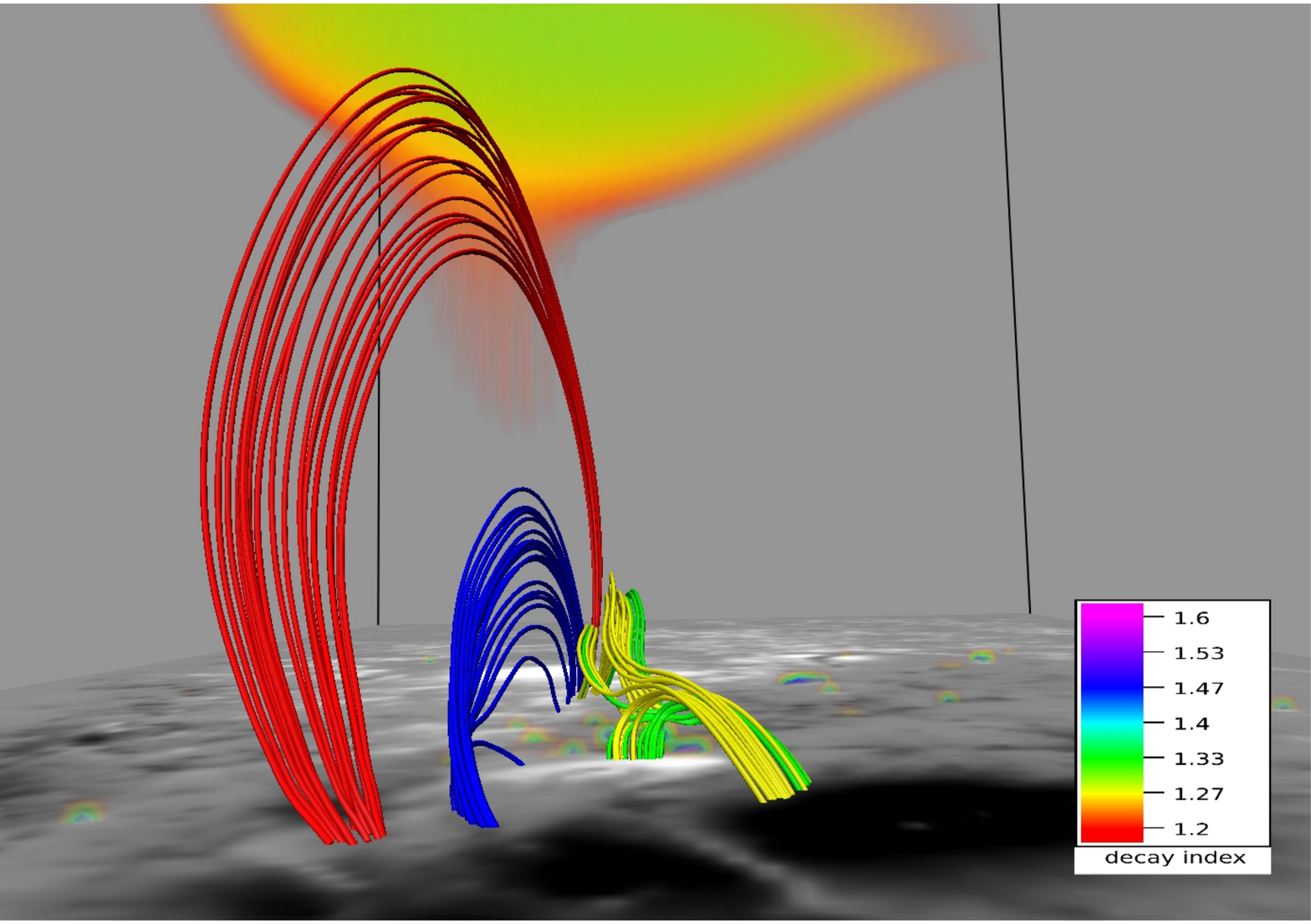}
    \caption{}
  \end{subfigure}
\quad
  \begin{subfigure}[]{0.45\textwidth}
    \centering
    \includegraphics[width=1\linewidth]{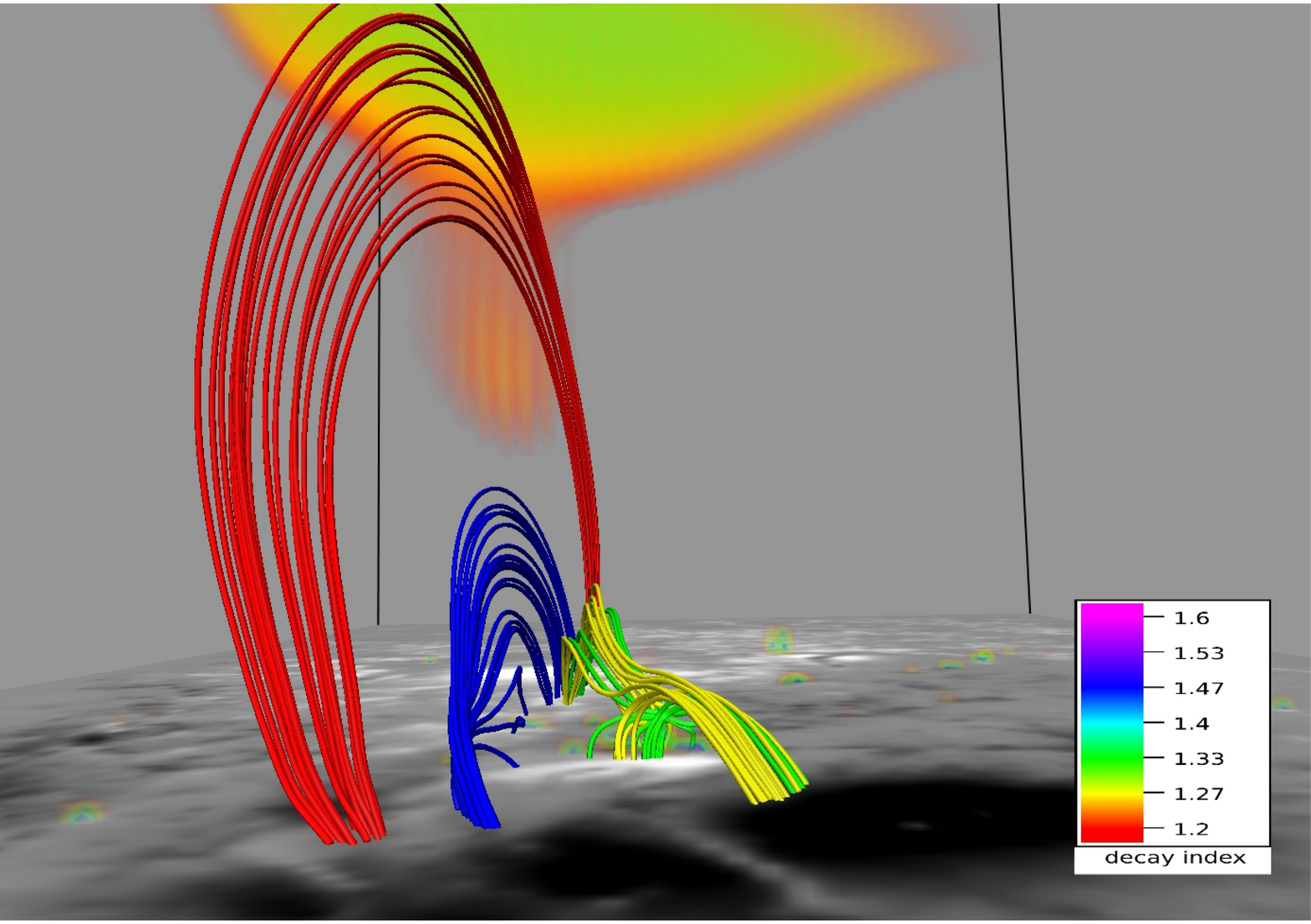}
    \caption{}
  \end{subfigure}
\quad
\begin{subfigure}[]{0.45\textwidth}
    \centering
    \includegraphics[width=1\linewidth]{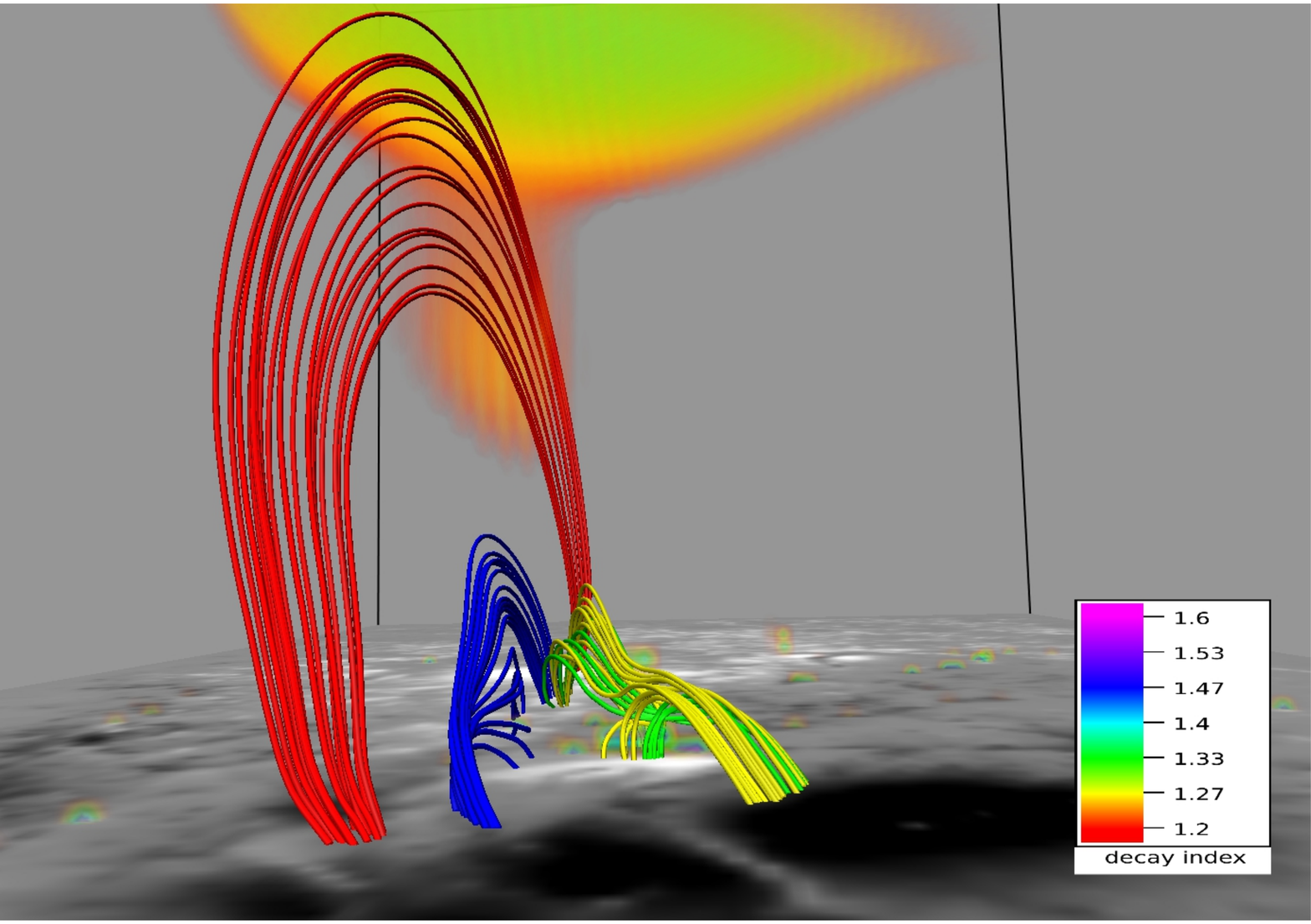}
    \caption{}
  \end{subfigure}
\quad
  \begin{subfigure}[]{0.45\textwidth}
    \centering
    \includegraphics[width=1\linewidth]{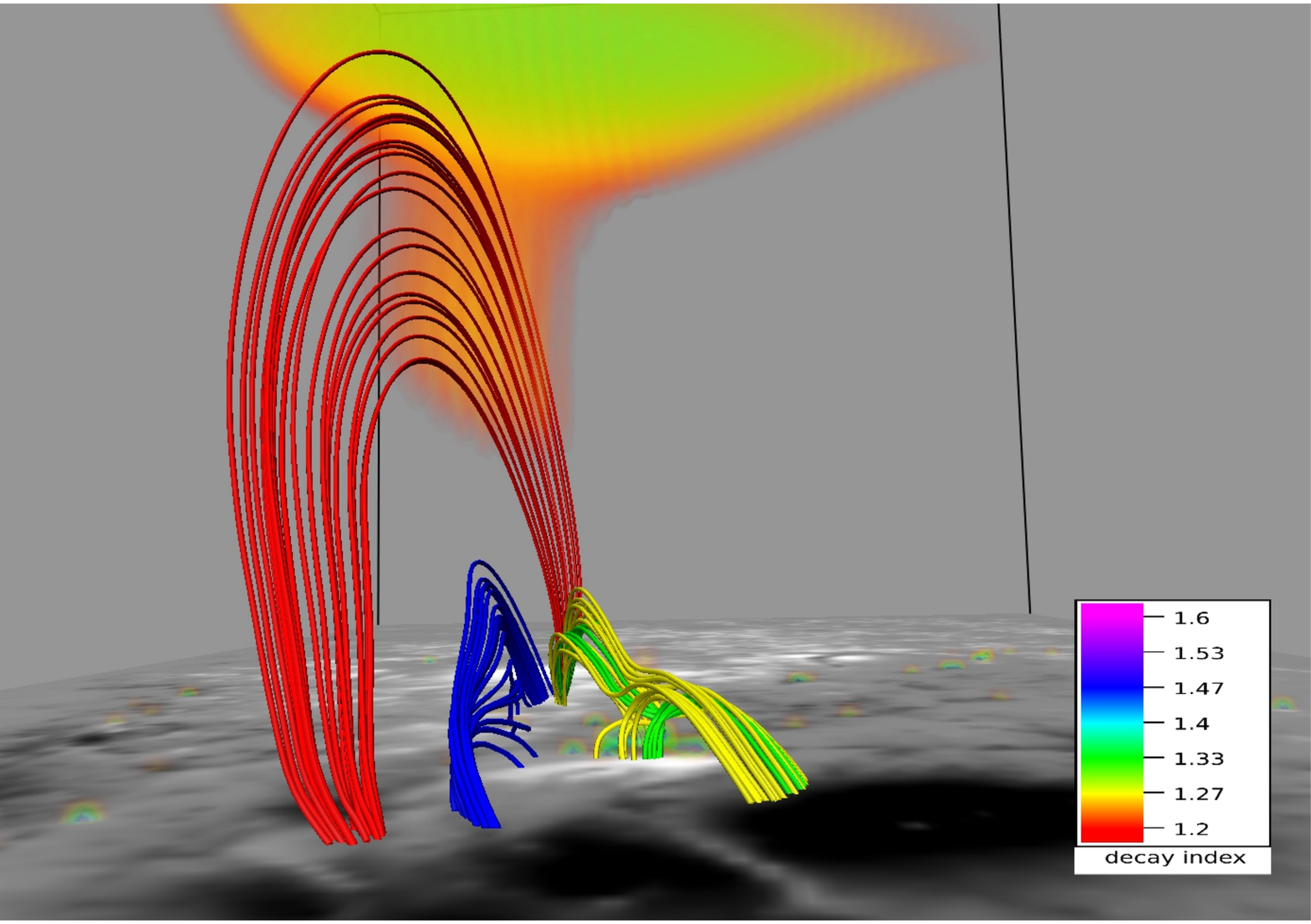}
    \caption{}
  \end{subfigure}
\caption{Front view of the evolution of the magnetic field lines previously shown in Figure \ref{f:recon-side}. In addition, we have shown the evolution of isosurfaces of decay index close to the critical value of 1.5 to explain the halt in the rise of the field lines. (An animation of this figure is available.)}
  \label{f:di_front}
\end{figure}

Figures \ref{f:recon-side} and \ref{f:di_front} depict MFL evolution in the neighborhood of the 3D null at two different viewing angles. The $B_z$ contours are 
plotted on the bottom boundary. Four sets of MFLs are highlighted. The fan and the spine of the null are made by the yellow MFLs whereas the red MFLs are 
overlying the null. The blue MFLs are located inside the dome whereas the arcade below the null is formed by the green MFLs. 
With evolution, the null and the constituent yellow MFLs do not sustain an appreciable ascent whereas the red MFLs expand significantly to a 
threshold height ($\approx 78$Mm), after which they contract. 
To explore the underlying physics, we note the arcade MFLs (in green) and the dome (yellow) constitute an X-type geometry cf. panel (b) of Figure \ref{f:3dnull}. As reconnection occurs at the X-type null, blue MFLs come out of the dome and overlays it. The consequent increase in local magnetic pressure pushes the red MFLs upward, resulting in their overall rise. Furthermore, the red MFLs get stretched as they rise and  at a  threshold  generate enough magnetic tension to stop additional upward motion. The threshold  corresponds to a critical value of $\approx 1.5$ for the decay index, where the decay index is defined as $\displaystyle{n = -\frac{\partial (\log|\mathbf{B}|)}{\partial (\log z)}}$ \citep{2006PhRvL..96y5002K}.
 Figure \ref{f:di_front} confirms, in their  maximal rise, the MFLs can only  attain $n \simeq 1.3-1.4$ which is in conformity with the confined nature of the X3.1 flare. Moreover, like \citep{2016ApJ...828...62J}, we also fail to identify a flux rope, which further agrees with the confined nature of the flare.

\begin{figure}[hp]
   \centering
   \begin{subfigure}[]{0.47\textwidth}
     \centering
     \includegraphics[width=1\linewidth]{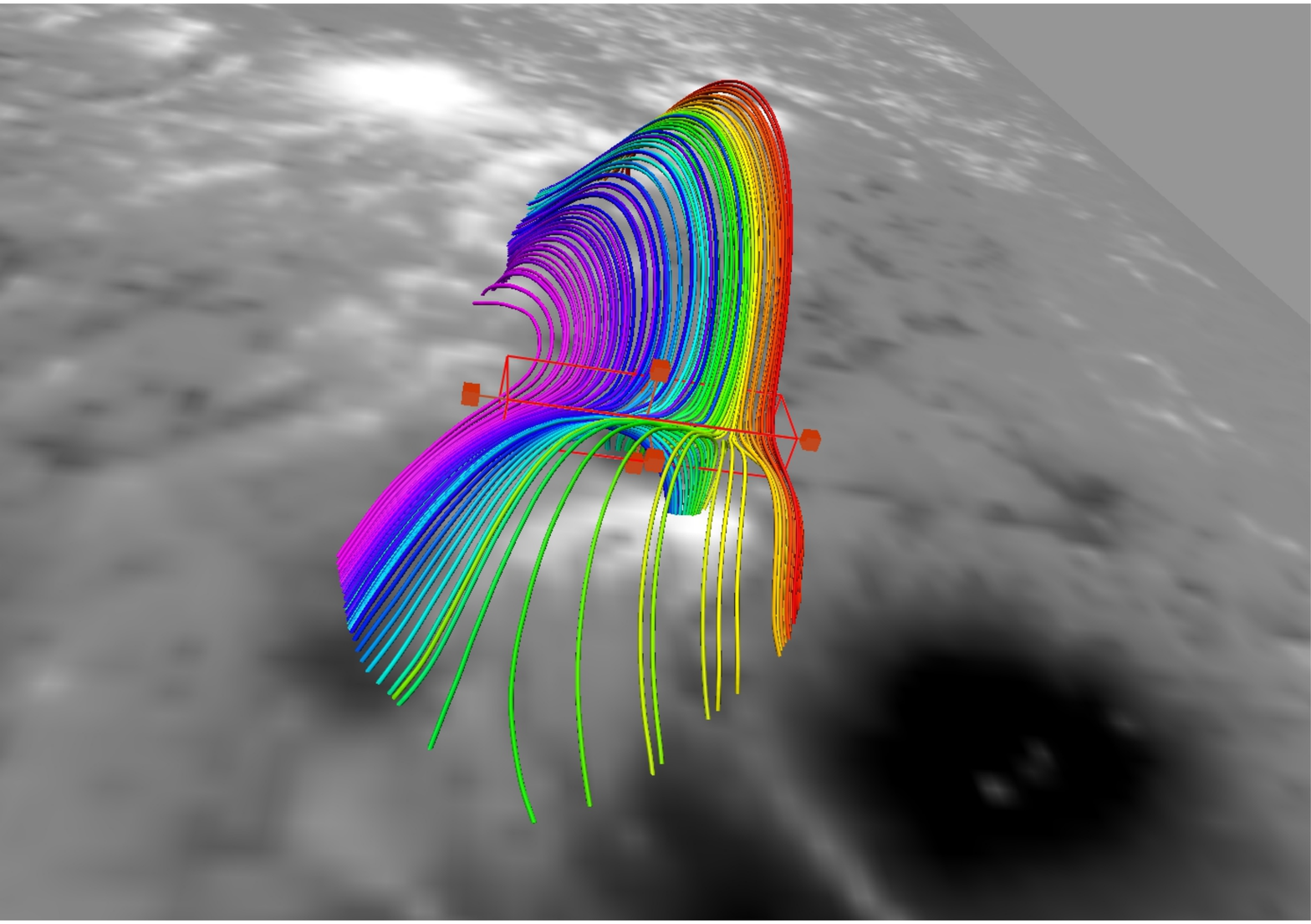}
     \caption{}
   \end{subfigure}
 \quad
   \begin{subfigure}[]{0.47\textwidth}
     \centering
     \includegraphics[width=1\linewidth]{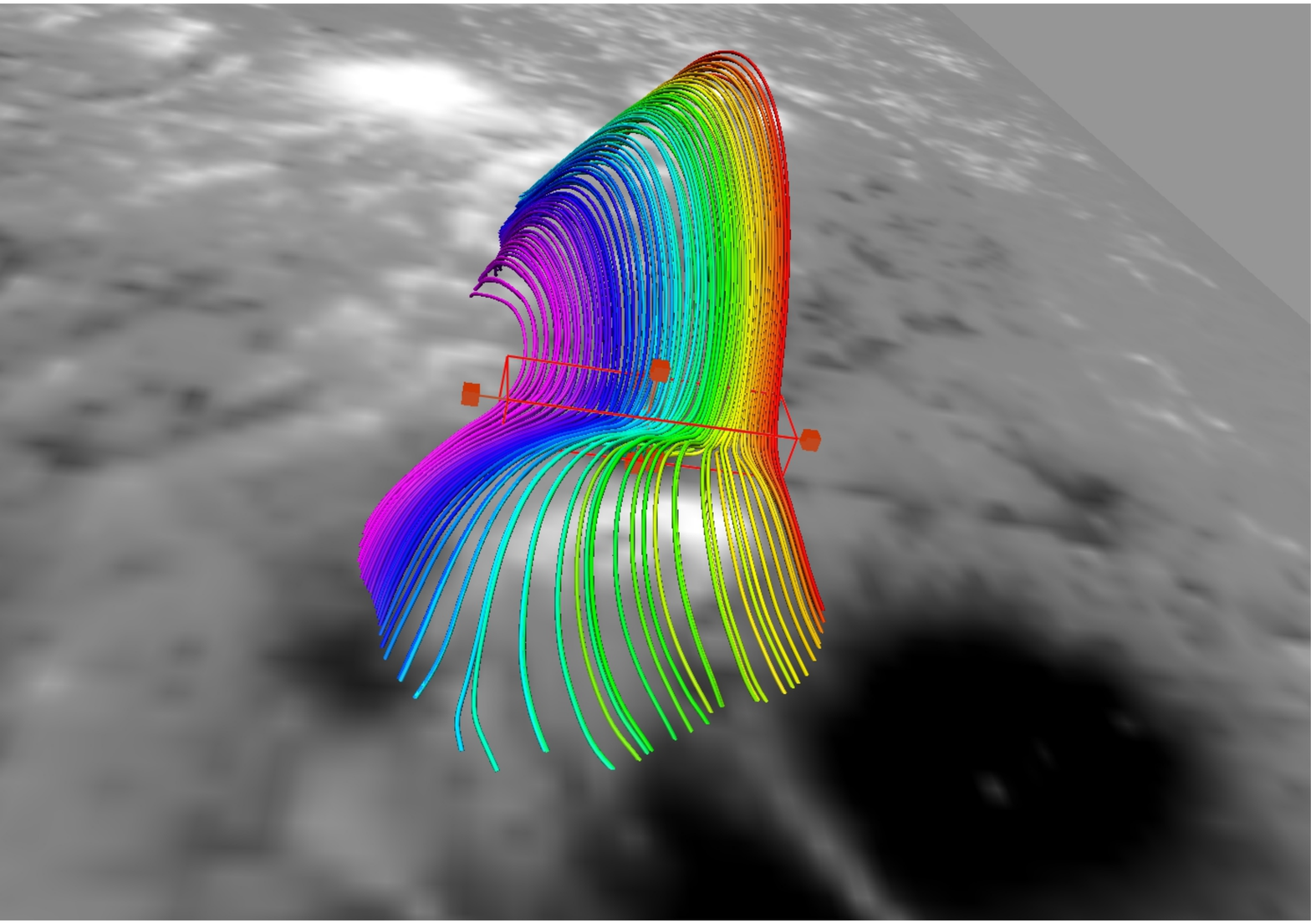}
     \caption{}
   \end{subfigure}
 \quad
 \begin{subfigure}[]{0.47\textwidth}
     \centering
     \includegraphics[width=1\linewidth]{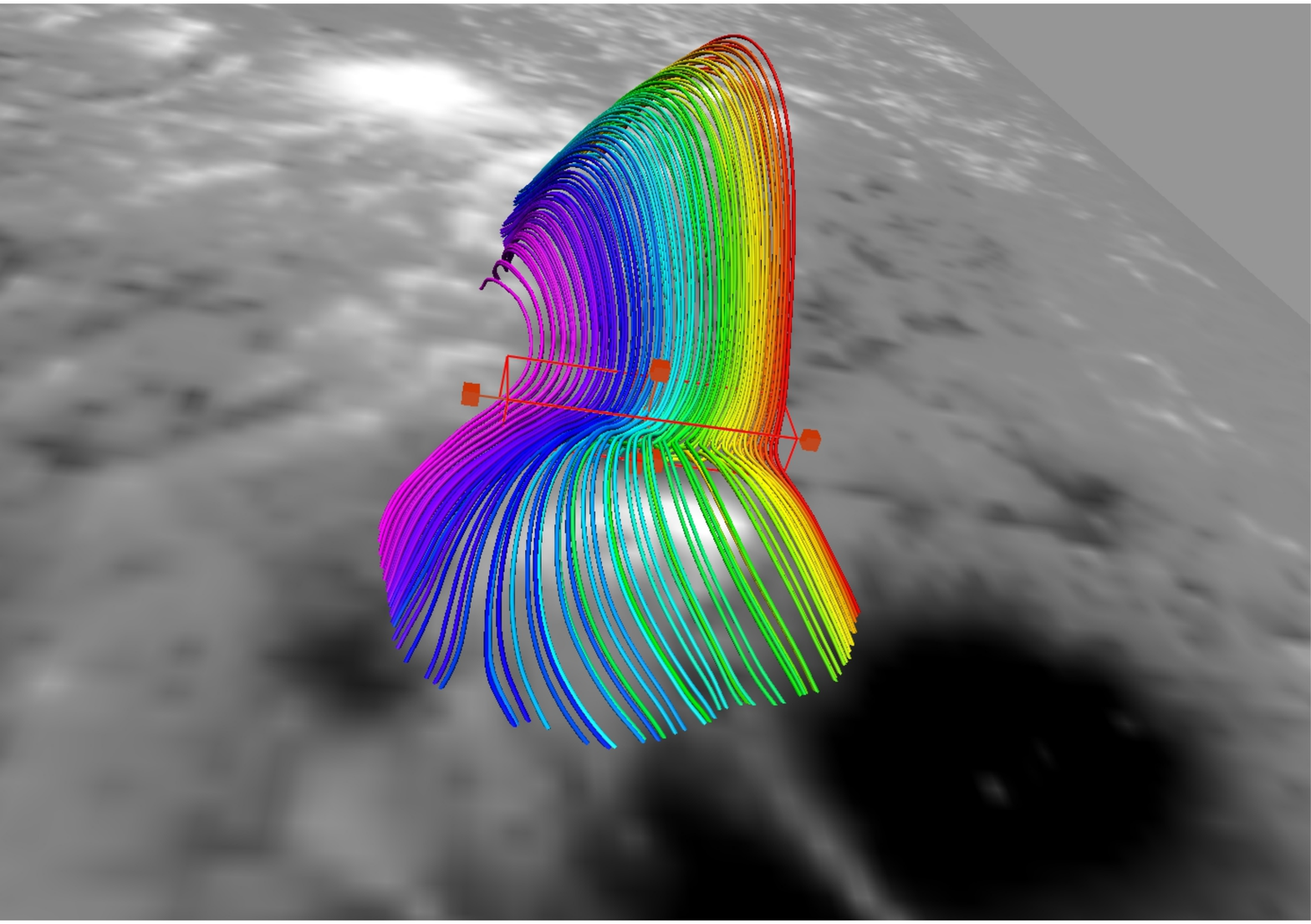}
     \caption{}
   \end{subfigure}
 \quad
   \begin{subfigure}[]{0.47\textwidth}
     \centering
     \includegraphics[width=1\linewidth]{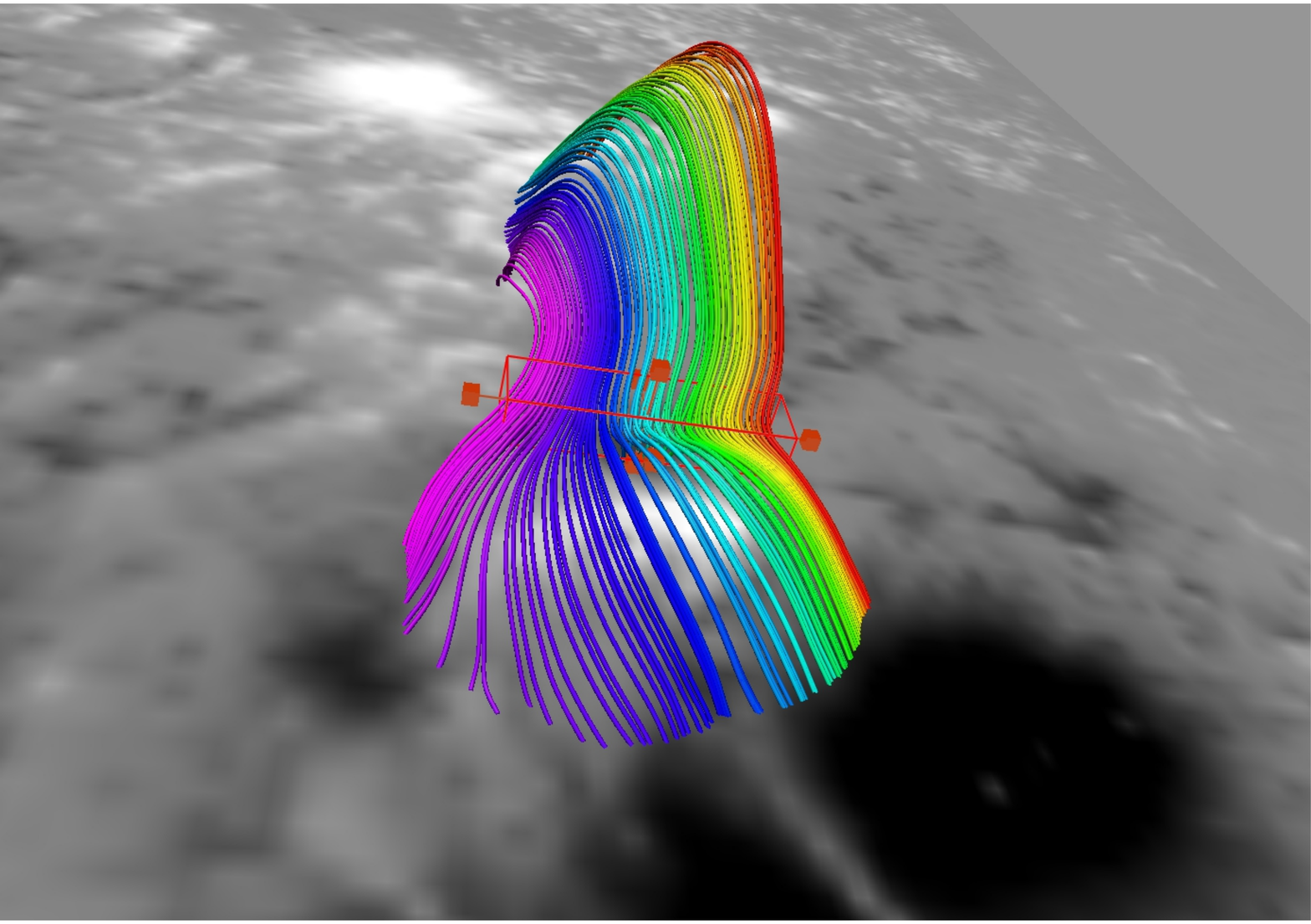}
     \caption{}
   \end{subfigure}
 \quad
 \begin{subfigure}[]{0.47\textwidth}
     \centering
     \includegraphics[width=1\linewidth]{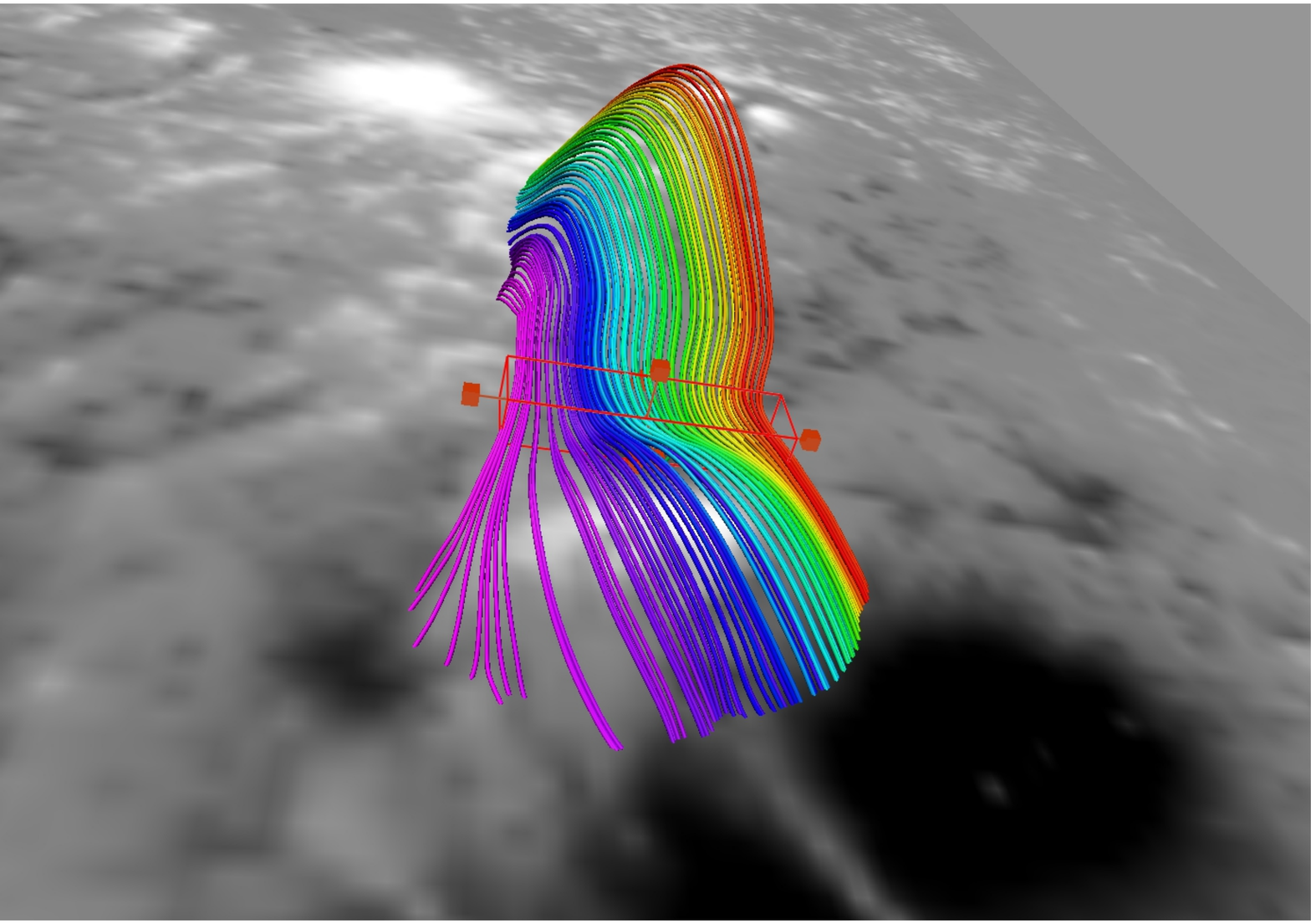}
     \caption{}
   \end{subfigure}
 \quad
   \begin{subfigure}[]{0.47\textwidth}
     \centering
     \includegraphics[width=1\linewidth]{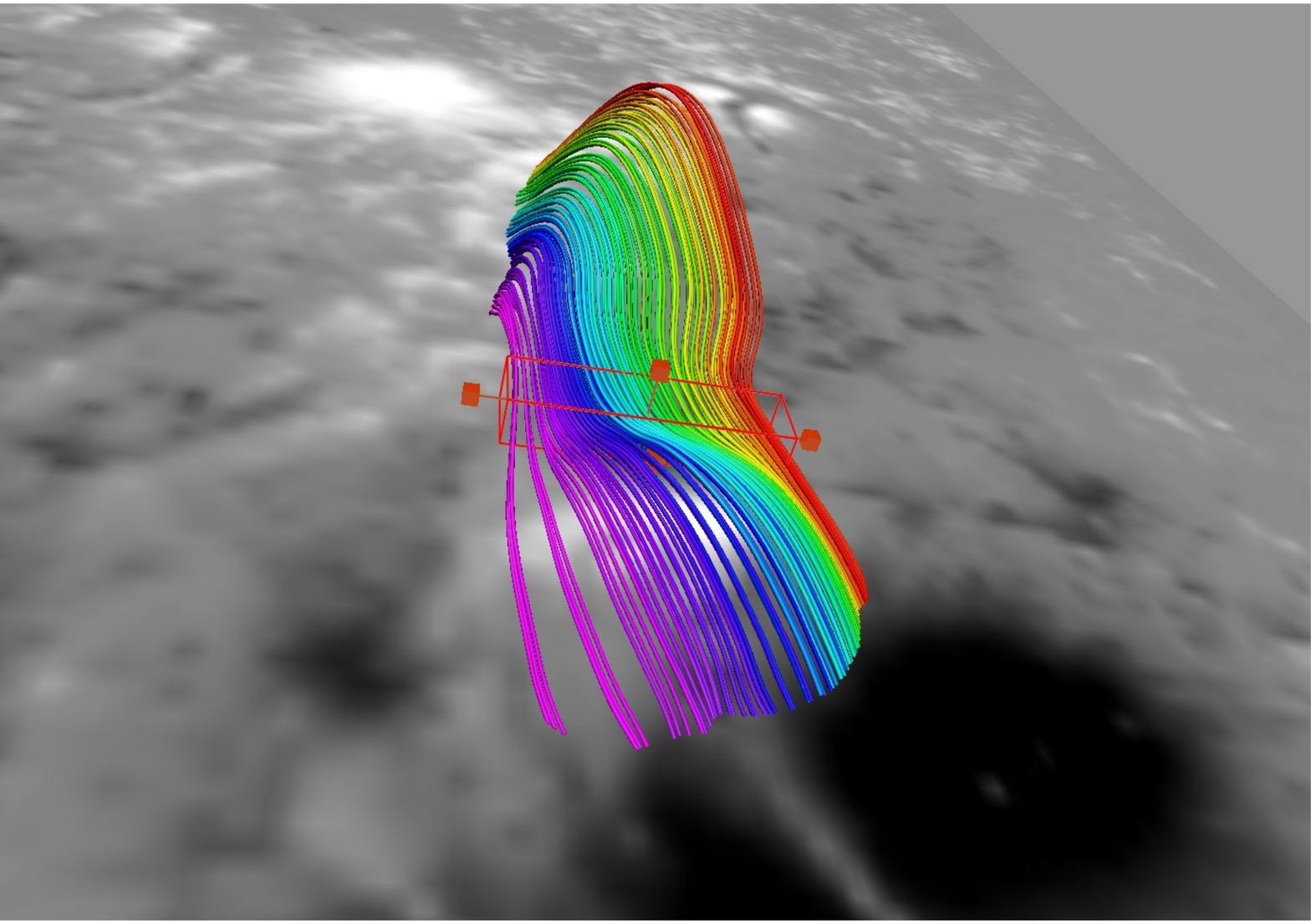}
     \caption{}
   \end{subfigure}
 \caption{Panels (a)-(f) spanning $t=$ 0, 80, 160, 240, 320 and 400,  illustrate rotation of the dome structure of the field lines constituiting the  3D null. The cuboidal rake in the figure shows the volume where the seed points are choosen. The field lines are color-coded with respect to their distance in the $y$ direction. This helps us to visualize the rotation of the field lines. (An animation of this figure is available.)}
   \label{f:null-rot}
 \end{figure}

\begin{figure}[hp]
   \centering
   \begin{subfigure}[]{0.47\textwidth}
     \centering
     \includegraphics[width=1\linewidth]{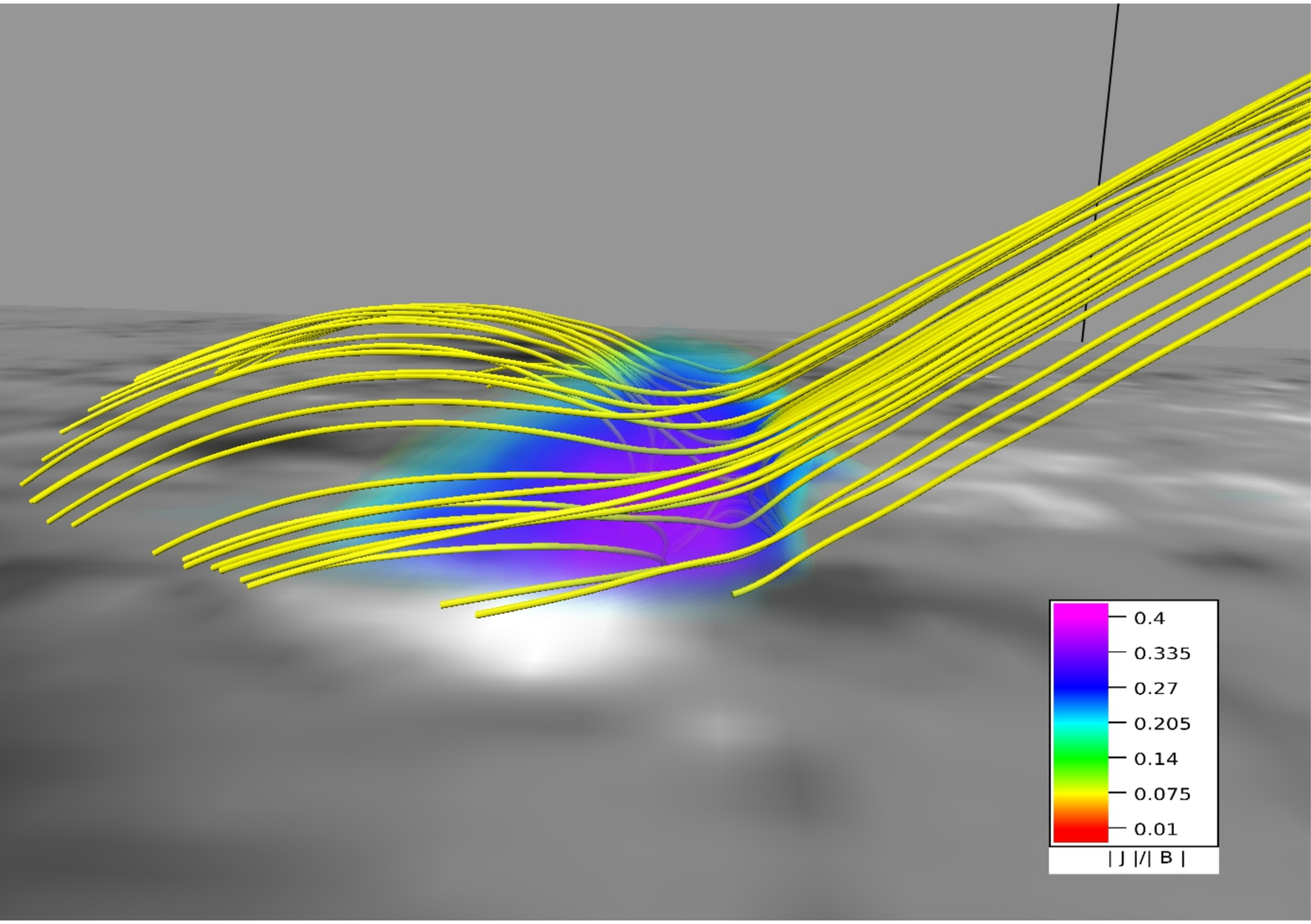}
     \caption{}
   \end{subfigure}
 \quad
   \begin{subfigure}[]{0.47\textwidth}
     \centering
     \includegraphics[width=1\linewidth]{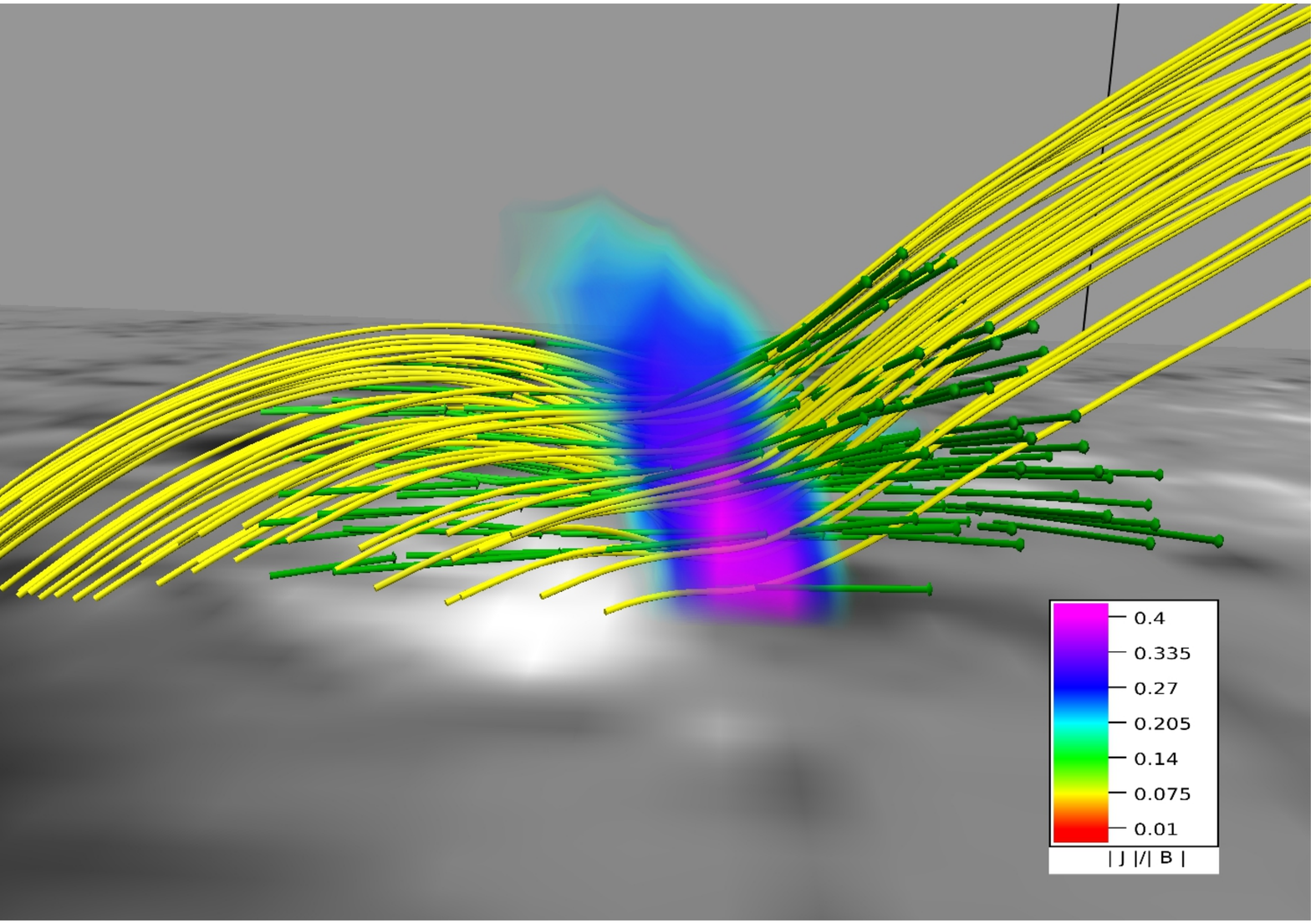}
     \caption{}
   \end{subfigure}
 \quad
 \begin{subfigure}[]{0.47\textwidth}
     \centering
     \includegraphics[width=1\linewidth]{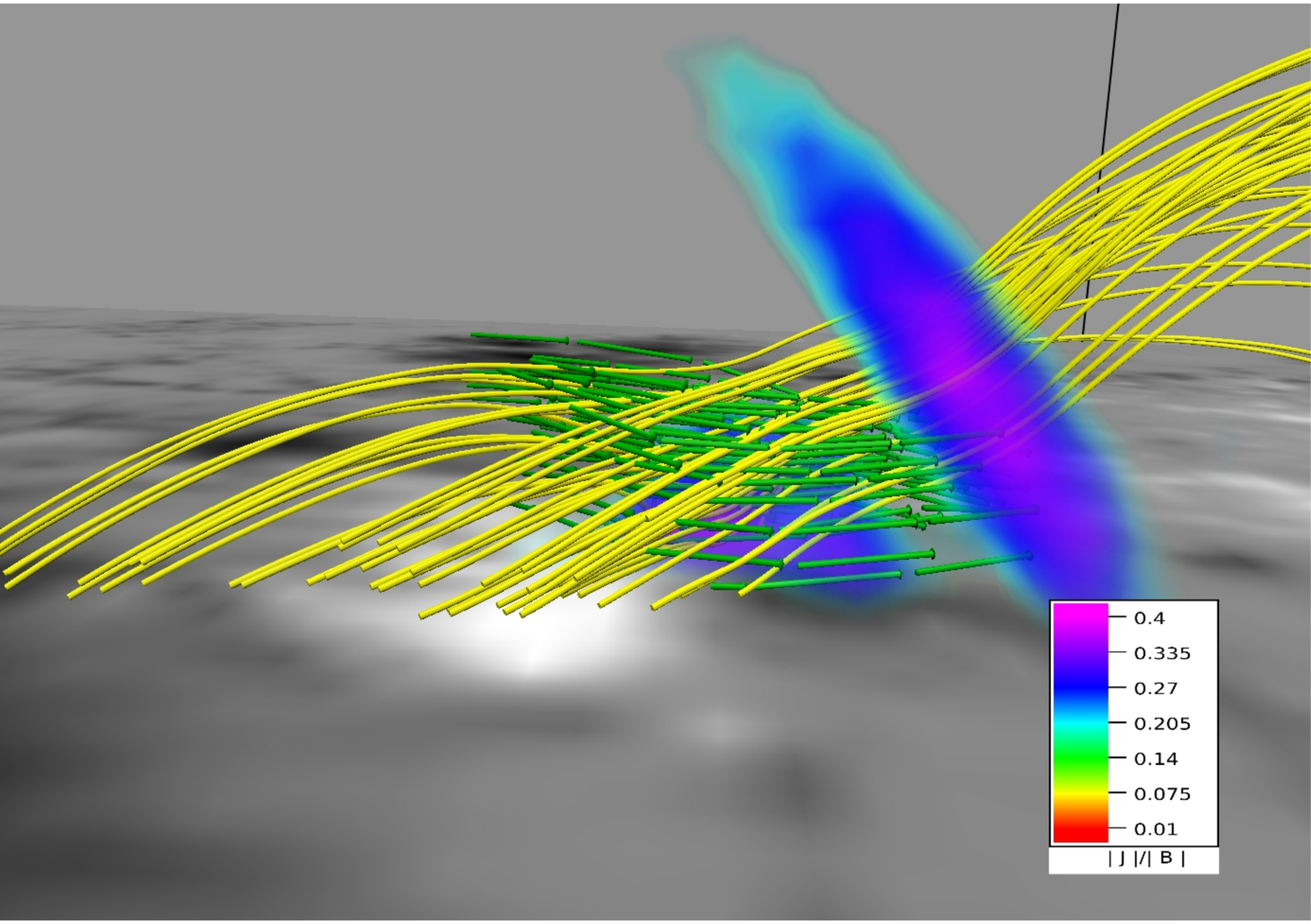}
     \caption{}
   \end{subfigure}
 \quad
   \begin{subfigure}[]{0.47\textwidth}
     \centering
     \includegraphics[width=1\linewidth]{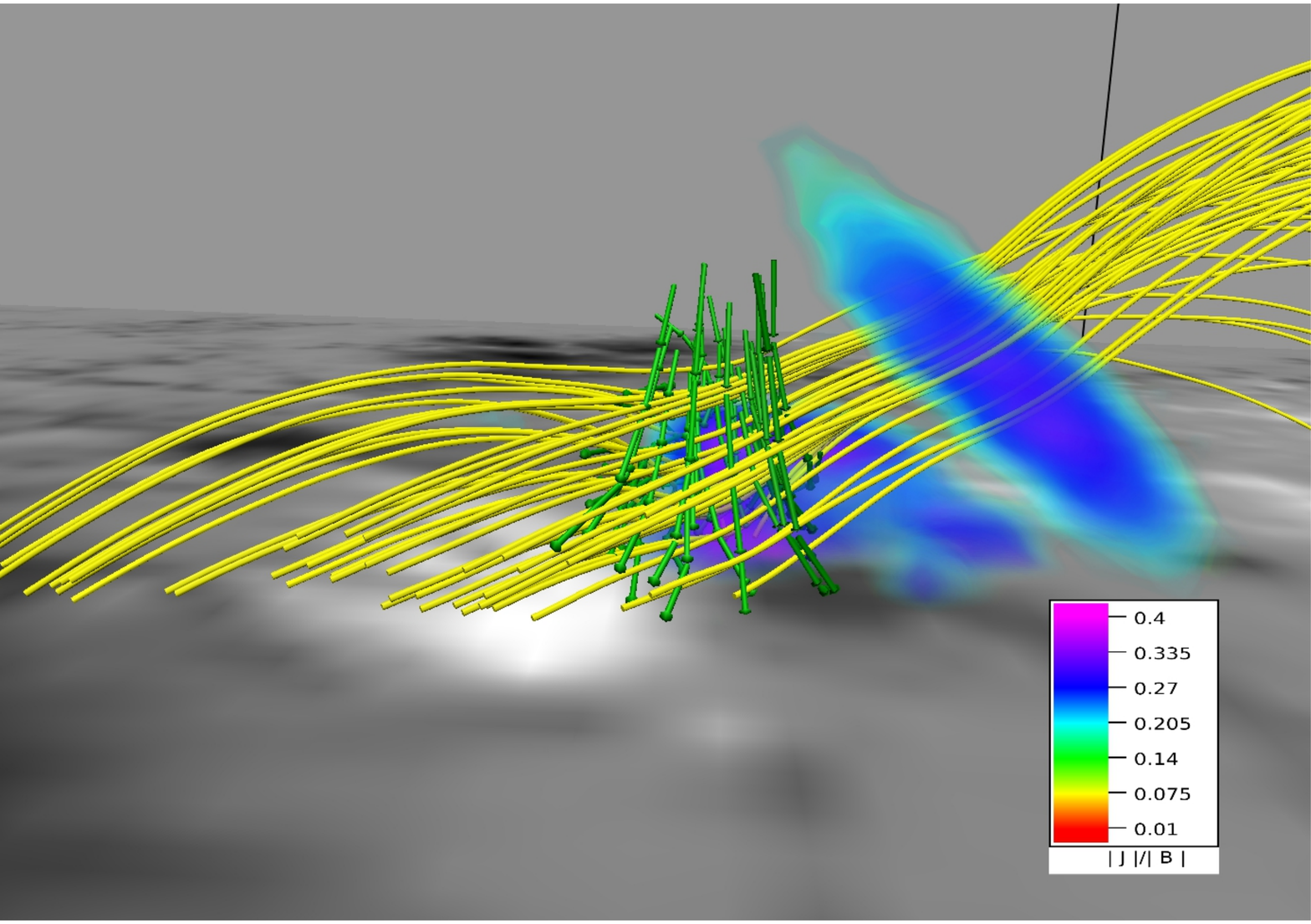}
     \caption{}
   \end{subfigure}
 \quad
 \begin{subfigure}[]{0.47\textwidth}
     \centering
     \includegraphics[width=1\linewidth]{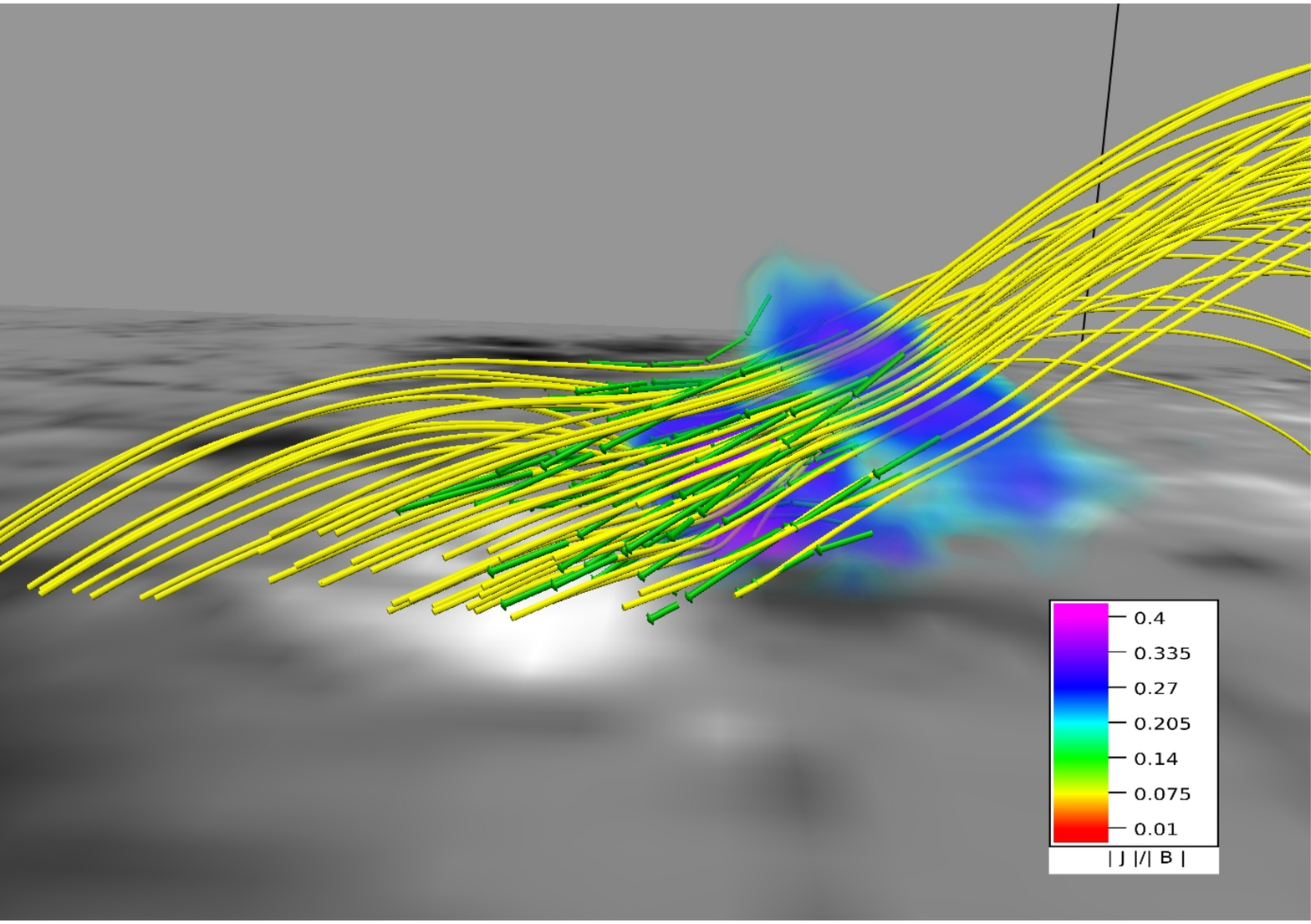}
     \caption{}
   \end{subfigure}
 \quad
   \begin{subfigure}[]{0.47\textwidth}
     \centering
     \includegraphics[width=1\linewidth]{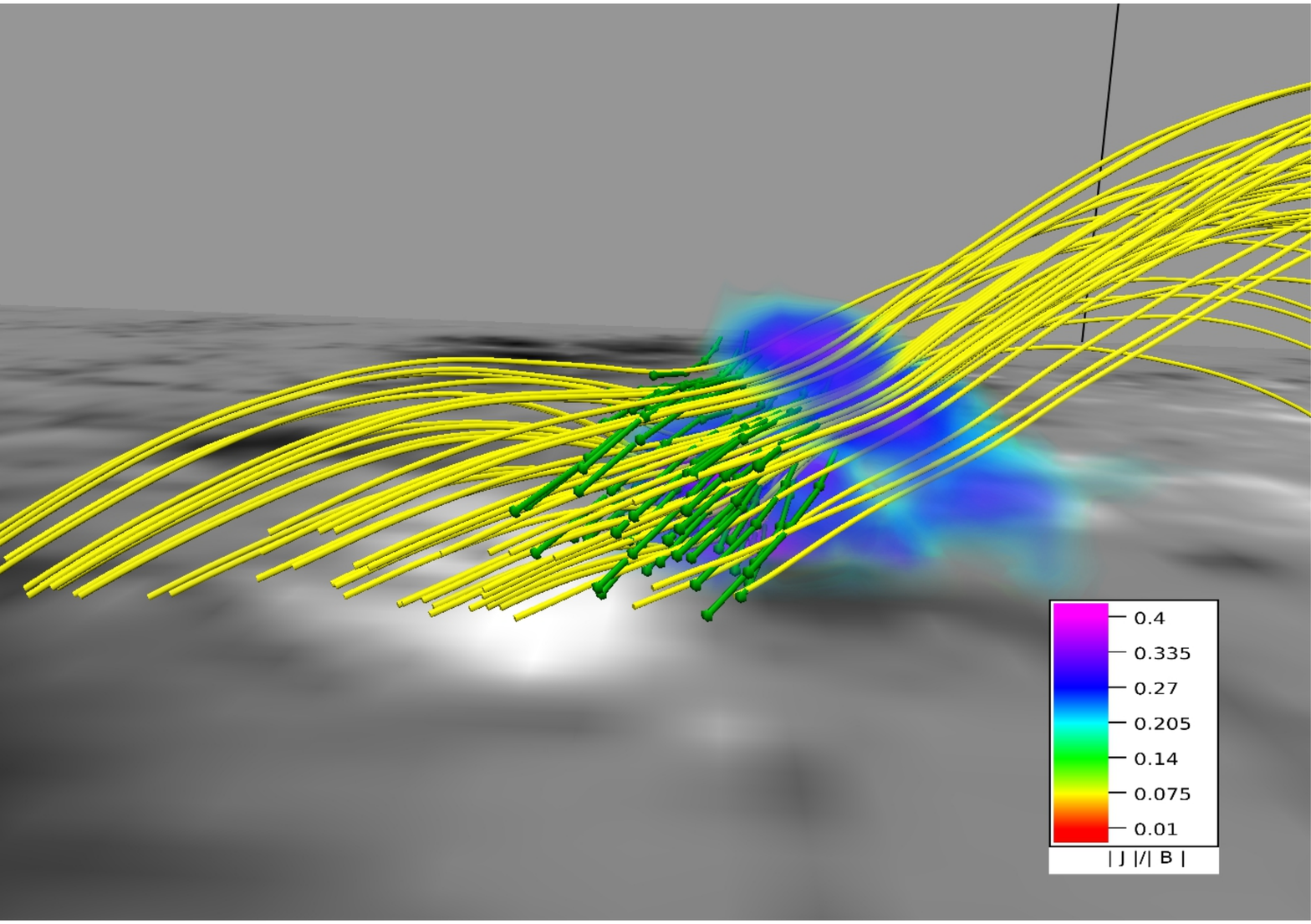}
     \caption{}
   \end{subfigure}
 \caption{Panels (a)-(f) spanning $t=$ 0, 200, 400, 600, 800 and 1000,  illustrate  the evolution of magnetic field lines (yellow), velocity field (green) and $|\mathbf{J}|/|\mathbf{B}|$ (An animation of this figure is available.)}
   \label{f:jb}
 \end{figure}

\begin{figure}[hp]
   \centering
   \begin{subfigure}[]{0.47\textwidth}
     \centering
     \includegraphics[width=1\linewidth]{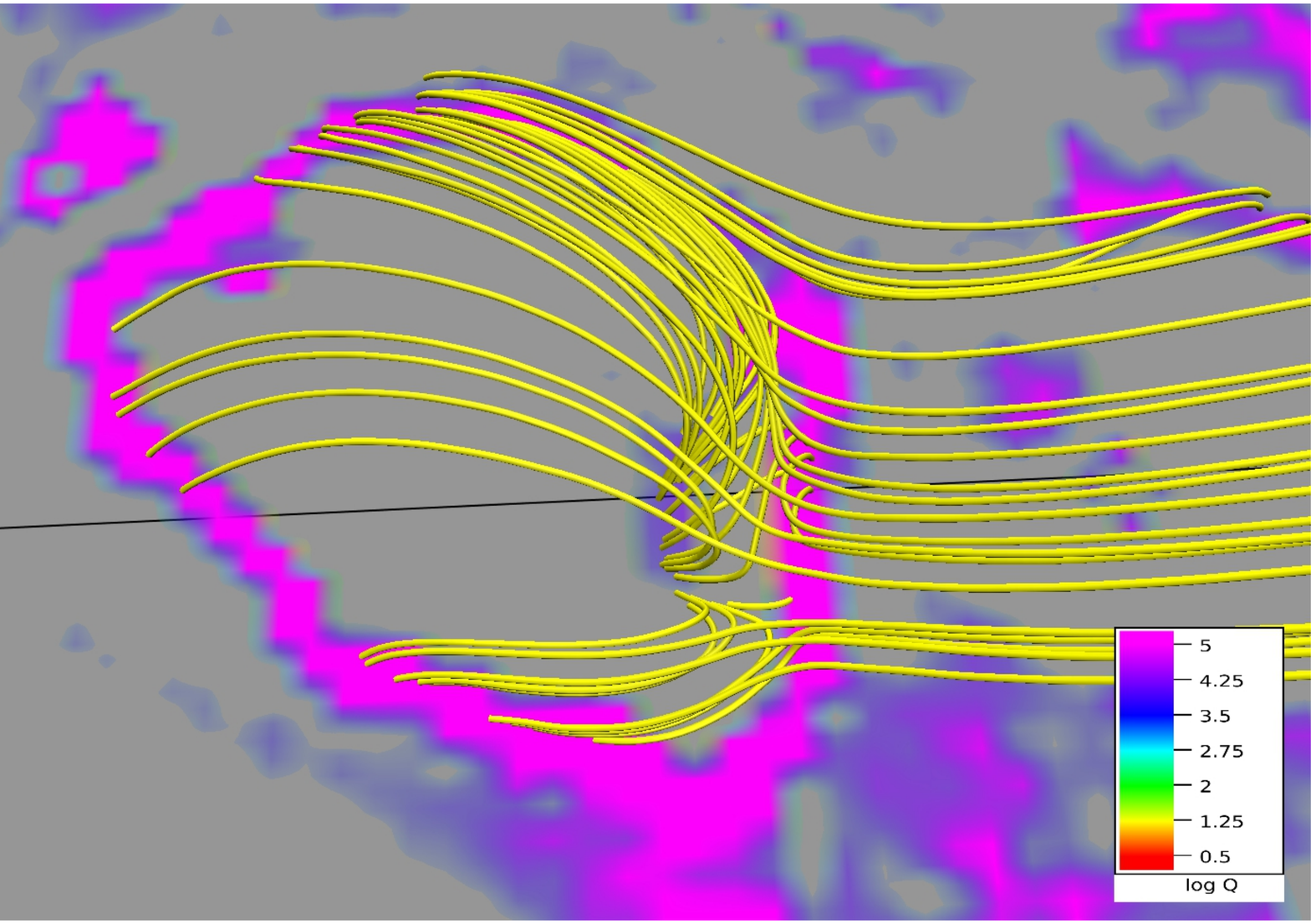}
     \caption{}
   \end{subfigure}
 \quad
   \begin{subfigure}[]{0.47\textwidth}
     \centering
     \includegraphics[width=1\linewidth]{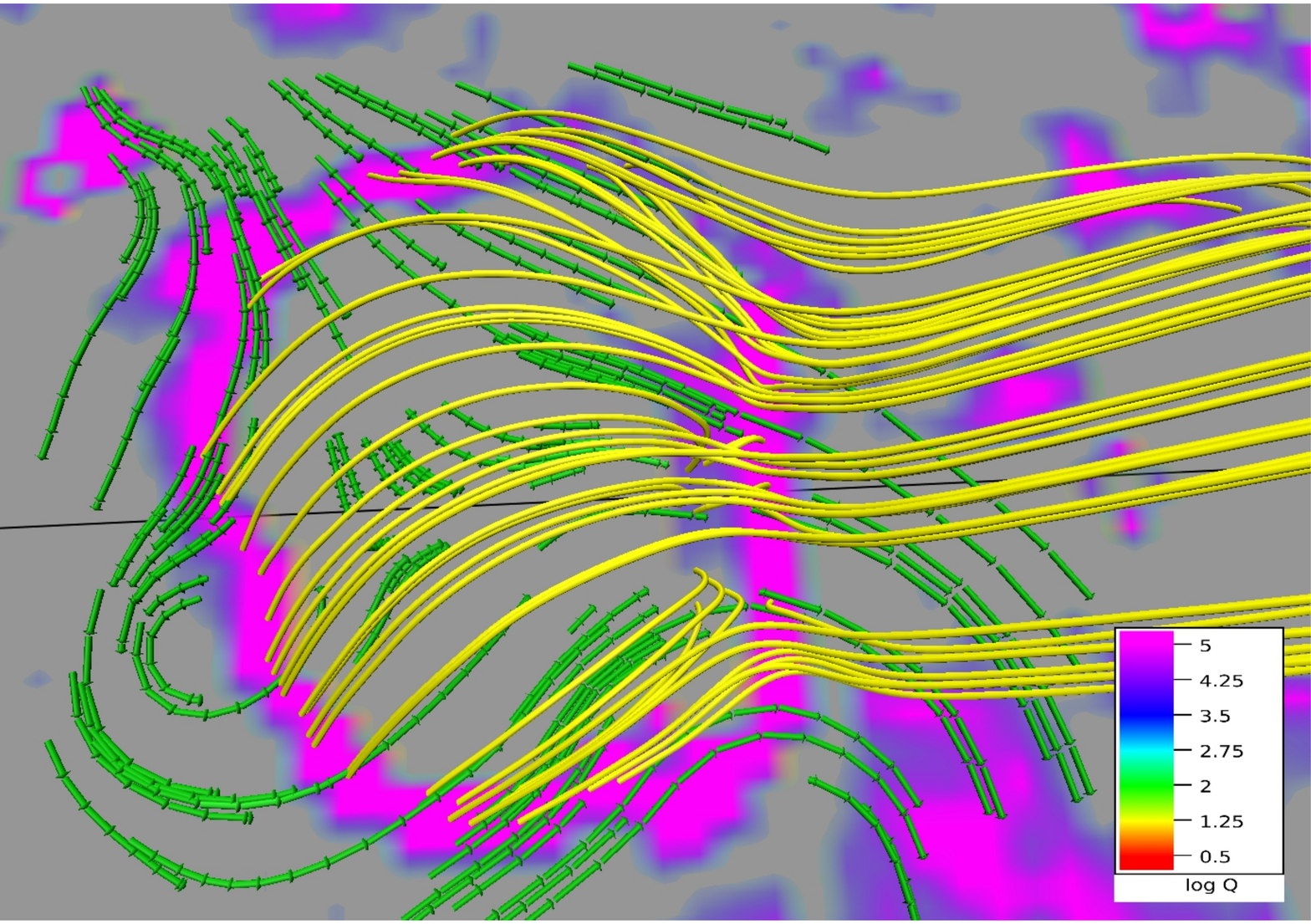}
     \caption{}
   \end{subfigure}
 \quad
 \begin{subfigure}[]{0.47\textwidth}
     \centering
     \includegraphics[width=1\linewidth]{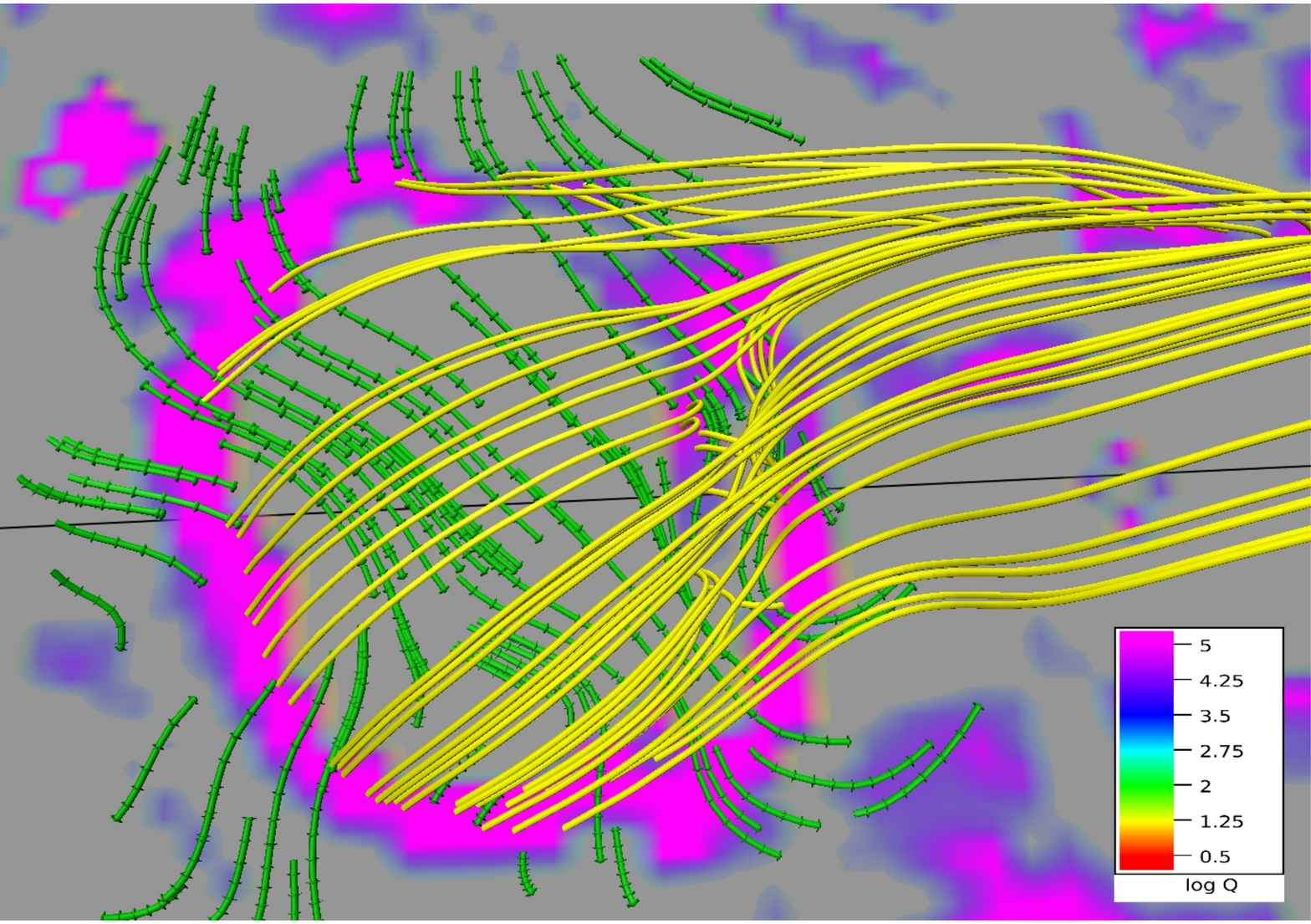}
     \caption{}
   \end{subfigure}
 \quad
   \begin{subfigure}[]{0.47\textwidth}
     \centering
     \includegraphics[width=1\linewidth]{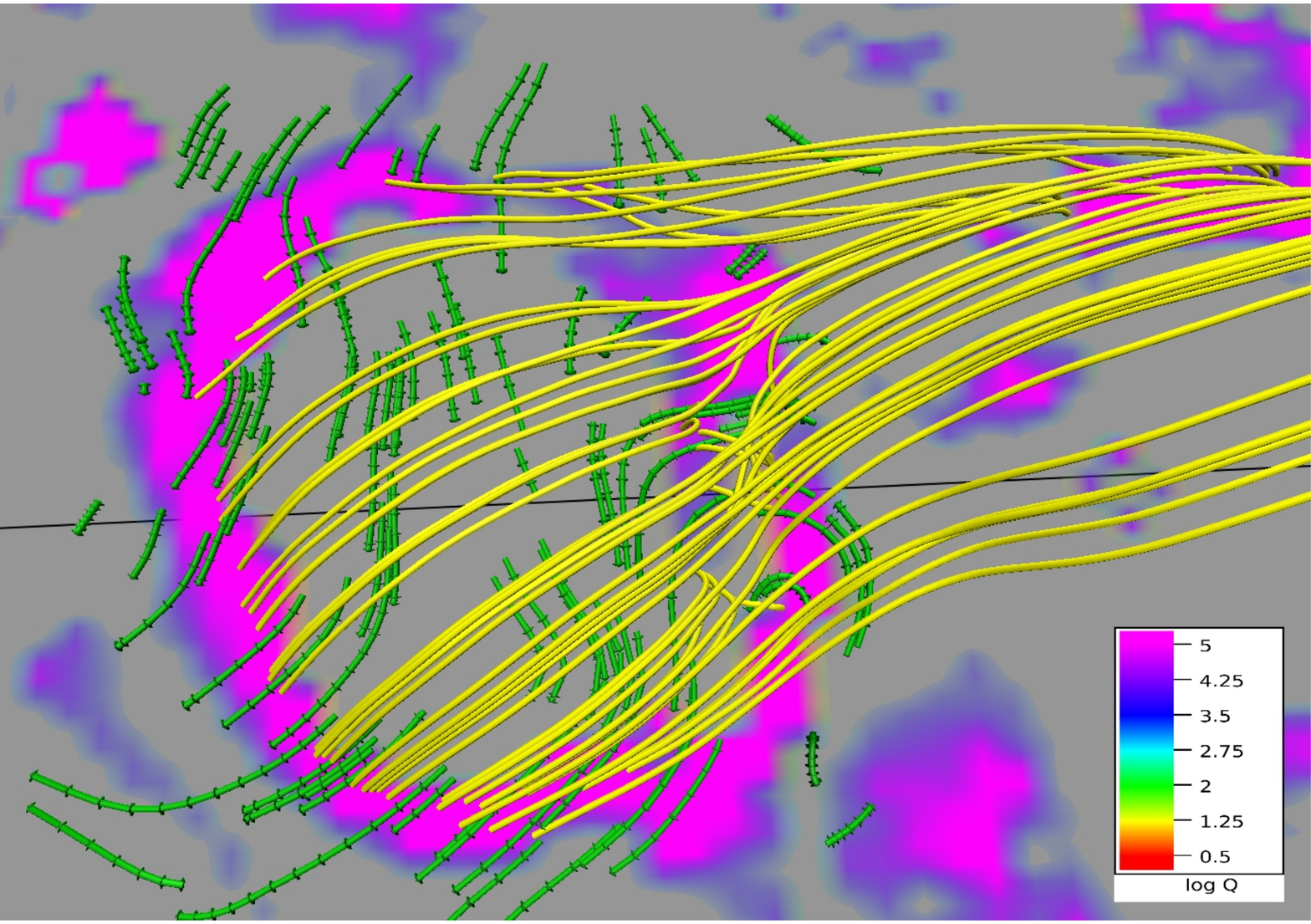}
     \caption{}
   \end{subfigure}
 \quad
 \begin{subfigure}[]{0.47\textwidth}
     \centering
     \includegraphics[width=1\linewidth]{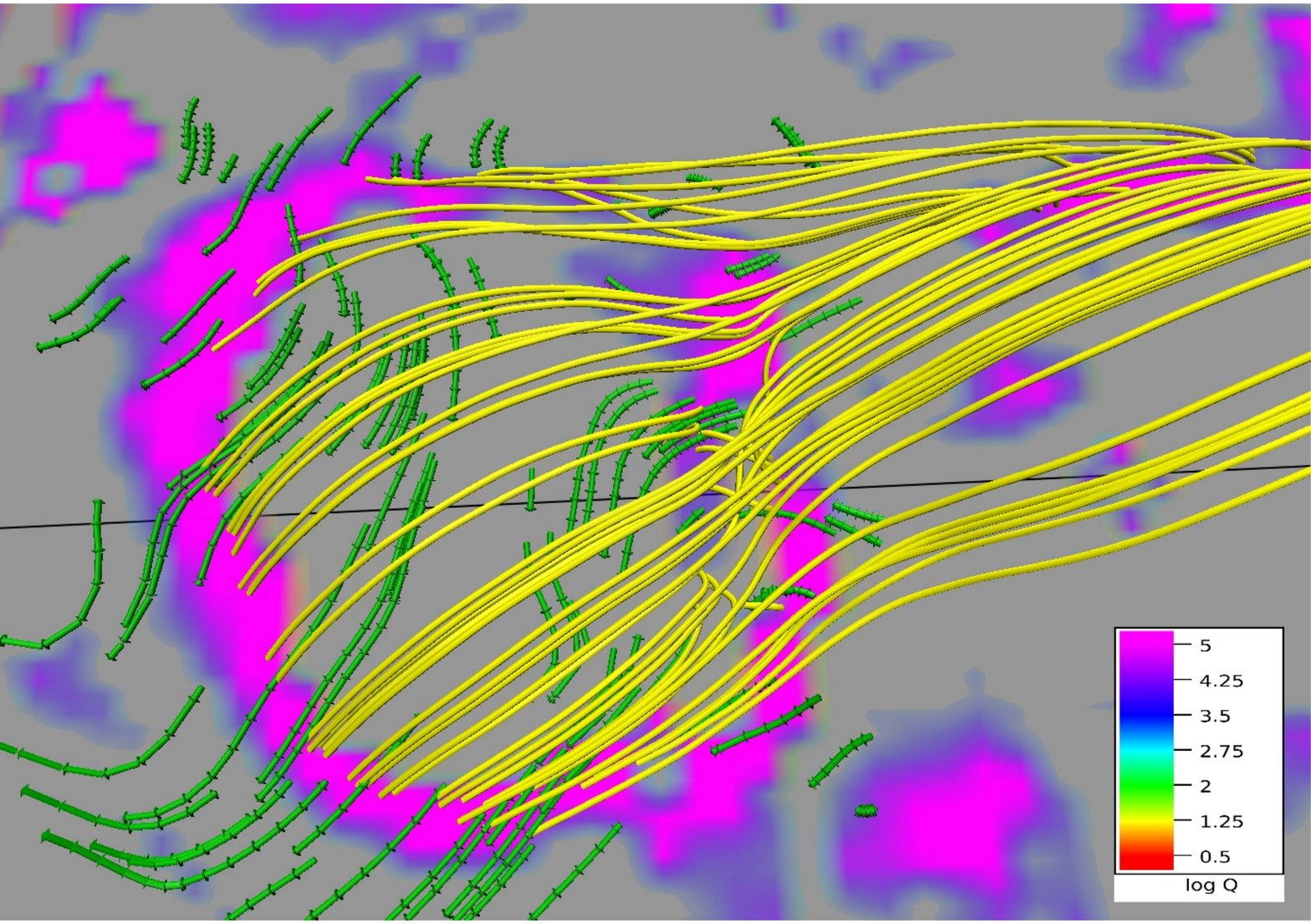}
     \caption{}
   \end{subfigure}
 \quad
   \begin{subfigure}[]{0.47\textwidth}
     \centering
     \includegraphics[width=1\linewidth]{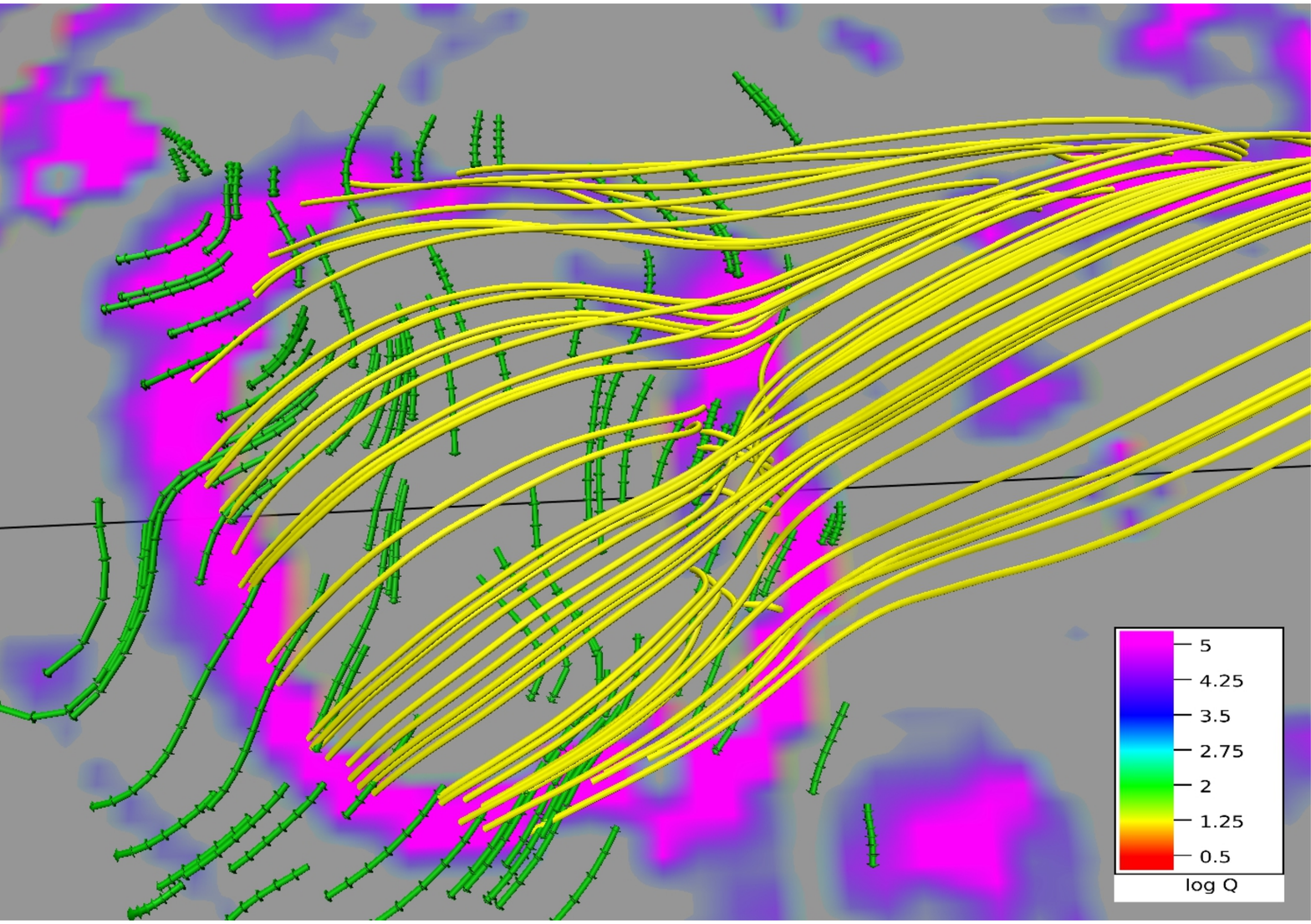}
     \caption{}
   \end{subfigure}
 \caption{Panels (a)-(f) spanning $t=$ 0, 200, 400, 600, 800 and 1000,  illustrate the slipping reconnections in the MFLs (shown in yellow) spanning the dome of the 3D null. The streamlines of the flow are shown in green. The bottom boundary shows contours of high values of $\log~ Q$. (An animation of this figure is available.)}
   \label{f:qsl}
 \end{figure}

The simulated 3D null appears to rotate with evolution (Figure \ref{f:null-rot}). 
For aiding visualization, the MFLs have been color-coded based on their distance along the $y$ axis. We have also shown the volume wherein the seed points of MFLs are located. When viewed from the top, an anti-clockwise rotation of the MFLs is quite prominent which matched very well with the similar dynamics seen in the AIA 131 \AA~ channel. This correspondence with observations gives more credibility to the simulation. The Figure \ref{f:jb} is also overlaid with streamlines (green) and $|\mathbf{J}|/|\mathbf{B}|$. Noticeable is the initial high value of $|\mathbf{J}|/|\mathbf{B}|$ near the null.
The value increases with time, becoming maximum at $t= 400$, decaying subsequently. The peaking of $|\mathbf{J}|/|\mathbf{B}|$ is indicative of magnetic reconnections occurring near the null. The resultant outflow is shown by the red streamlines. For further investigation,  Figure \ref{f:qsl} plots the Q-map where the squashing factor Q is calculated by following \citet{1996A&A...308..643D,2016ApJ...818..148L} and ascertains the dome to have high gradient of magnetic connectivity which results in slipping reconnections \citep {2007Sci...318.1588A}. The subsequent change in magnetic connectivity manifests as the seeming MFL rotation. For validation, we note the co-located flow (in green) is 
not along the  rotation and hence, cannot cause it.


\begin{figure}[hp]
  \centering
  \begin{subfigure}[]{0.75\textwidth}
    \centering
    \includegraphics[width=1\linewidth]{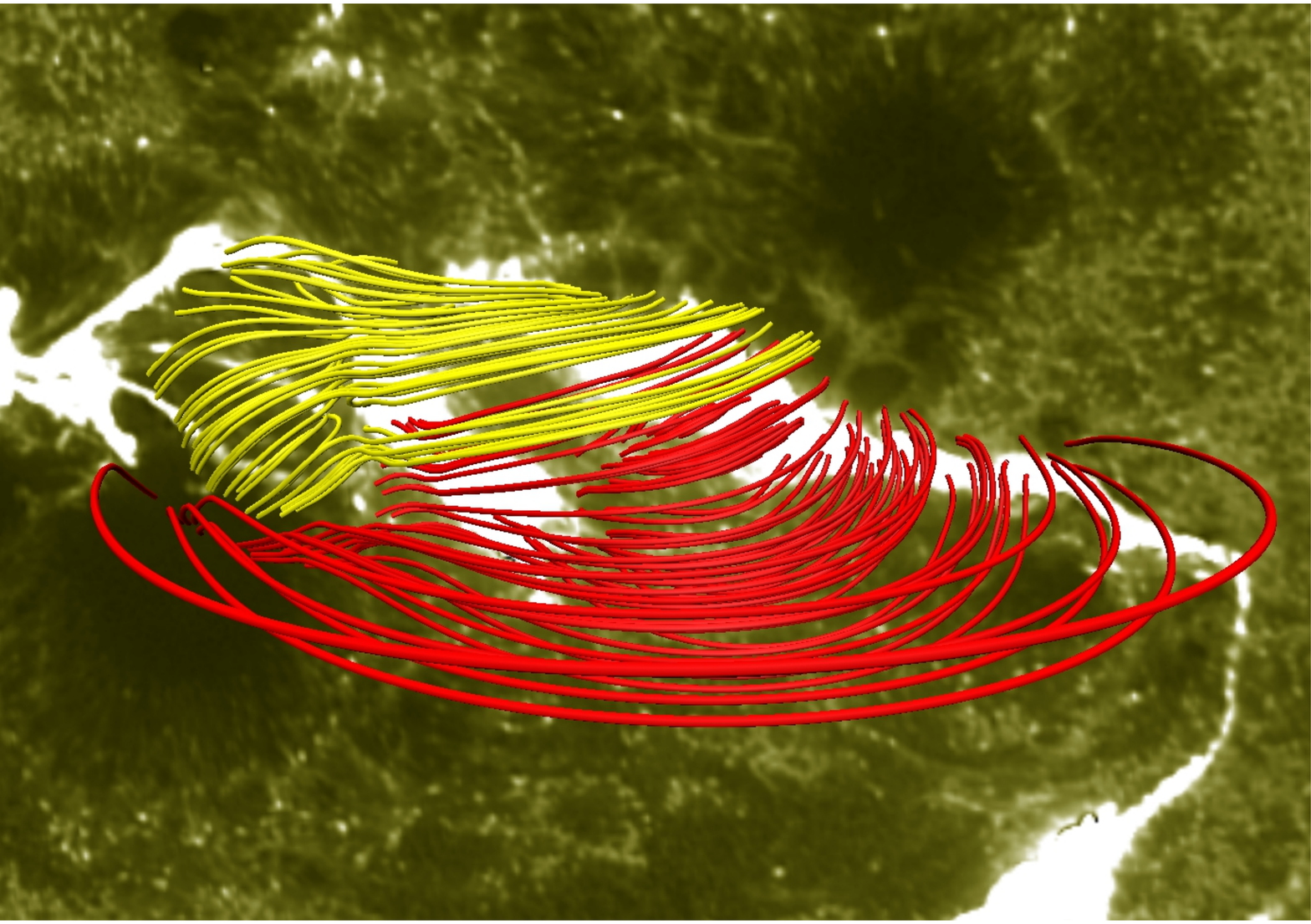}
    \caption{}
  \end{subfigure}
  \begin{subfigure}[]{0.75\textwidth}
    \centering
    \includegraphics[width=1\linewidth]{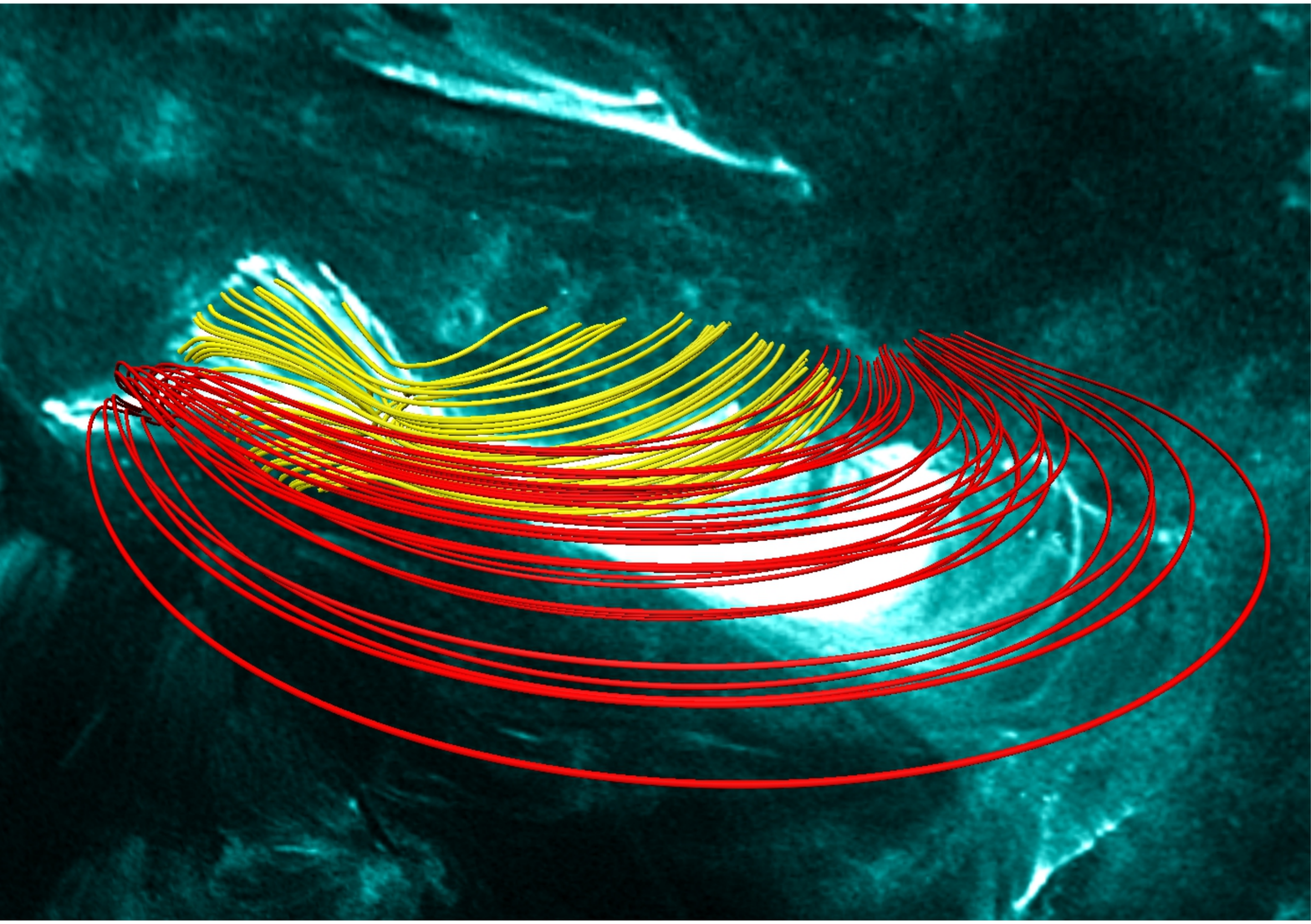}
    \caption{}
  \end{subfigure}
  \caption{The AIA 1600 \AA~ and 131 \AA~ channel images as depicted in Figure \ref{f:aia} are overlaid with the relevant magnetic field lines. Important is the almost exact match of footpoints with the location of the brightening.}
  \label{f:aia-match}
\end{figure}

In panels (a) and (b) of Figure \ref{f:aia-match}, we overlay  intensity structures in  wavelengths  1600 \AA~  at 21:25 UT and 131 \AA~ at 20:50 UT with
 corresponding MFLs. Importantly, the almost exact match of the footpoints for both wavelengths with brightenings not only establishes the importance of the 3D null in the circular flare ribbon but also being in agreement with the contemporary understanding, validates the effectiveness of the NFFF extrapolation in constructing a valid coronal field model.

\section{Summary and conclusions}

The paper presents simulated dynamics of AR12192 from 20:48 UT. The plasma is idealized to have perfect electrical conductivity while being viscid, thermally homogeneous and incompressible. The simulations are initialized with magnetic field lines extrapolated from SDO/HMI vector magnetograms using a new technique which employs a model where the corona is not strictly force-free and has some Lorentz force. Nevertheless, the Lorentz force decreases rapidly with height making the corona to be force-free in an asymptotic limit---agreeing with the standard scenario of the coronal field.  Advantageously, this non-force-free-field extrapolation model self-consistently initialize the coronal dynamics without requiring prescribed plasma flows which are somewhat custom-made. 

The extrapolated magnetic field is found to have a 3D null located approximately at a height of 3 Mm from the photosphere and has clearly distinguishable spine and a dome shaped fan. Importantly, a magnetic arcade is found  to be located within the dome and making an X-type null with it.  A Q-map of the initial field identifies the dome with a region where the gradient of the field line connectivity is large.

The simulation focuses on a circular brightening recorded in 1600 \AA~ channel at around 21:21 UT. The absence of any flux emergence in the window of 19:00 UT-24:00 UT allows the vertical field at the bottom boundary to be assumed as line-tied. 
To optimize the computation cost, the simulation is performed on $256\times128\times128$ grids along the $x$, $y$ and, $z$ respectively, resolving a physical domain of $360 \times 180 \times 180$ Mm$^3$.
It is  initiated not by a prescribed flow, but by  the initial Lorentz force which onset the evolution autonomously. Subsequently, the favorable forces bring non-parallel field lines in close proximity of the 3D 
null, which ultimately leads to unresolved scales. In the spirit of ILES, the MPDATA then generates locally adaptive residual dissipation to regularize the underresolved scales with simulated MRs. Further, the MRs are found to be consistent with the idea of slipping reconnection, standard at a 3D null, and imparts a sense of rotation to the footpoints of the dome. Such a rotation is also observed in the channel \AA, corroborating the observed circular brightening to be caused by MRs at the 3D null.

Magnetic reconnections also occur at the X-type null formed by the MFLs belonging to the arcade and the spine. Interestingly, reconnections enable MFLs contained within the dome to come out of it and overlay the spine. The consequent increase in magnetic pressure raises  the overlying MFLs further up and in principle, can cause the X3.1 flare observed at 21:15 UT. The flare was confined in nature and resulted in no CMEs. In the simulation, the MFLs are found to never reach a height where the decay index becomes more the critical value required for the torus instability to set in, confirming further the efficacy of the simulation in replicating the observation.


\acknowledgements 

\noindent {\bf Acknowledgements:} The simulations are performed using the 100 TF cluster Vikram-100 at the Physical Research Laboratory, India. We acknowledge the use of the visualization software VAPOR (www.vapor.ucar.edu) for generating relevant graphics. Data and images are courtesy of NASA/SDO and the HMI and AIA science teams. SDO/HMI is a joint effort of many teams and individuals to whom we are greatly indebted for providing the data. QH acknowledges partial support of NASA grant 80NSSC17K0016 and NSF award AGS-1650854. The authors are thankful to Dr. P. K. Smolarkiewicz for his support. We are also thankful to the anonymous referee for providing insightful suggestions toward the overall betterment of this paper.

\bibliography{ms}
\end{document}